\documentclass{aa} 

\usepackage{graphicx}
\usepackage[varg]{txfonts}
\usepackage{natbib}
\usepackage{lscape}
\usepackage[font={small,rm}]{caption}
\usepackage{array}
\usepackage{bm}
\usepackage{color}

\begin{document} 
    \title{Mid-IR spectra of Pre-Main Sequence Herbig stars:\\ an explanation for the non-detections of water lines}
\authorrunning{Antonellini et al.}
   \subtitle{}

\author{Antonellini$^{1}$, S. \and Kamp, I.$^{1}$ \and Lahuis, F.$^{2}$ \and Woitke, P.$^{3}$ \and Thi, W.-F.$^{4,5}$ \and Meijerink, R.$^{6}$ \and Aresu, G.$^{1,7}$ \and Spaans, M.$^{1}$  \and G\"{u}del, M.$^{8}$ 
\and Liebhart, A.$^{8}$}

	   \institute{Kapteyn Astronomical Institute,
		      Postbus 800, 9700 AV Groningen, The Netherlands\\
		      \email{antonellini@astro.rug.nl}\and
		      SRON Netherlands Institute for Space Research, P.O. Box 800, 9700 AV Groningen, The Netherlands\and
		      St. Andrews University, School of Physics and Astronomy\and
		      Univ. Grenoble Alpes, IPAG, F-38000 Grenoble, France CNRS, IPAG, F-38000 Grenoble, France\and
		      Max-Planck-Institut f$\rm\ddot{u}$r extraterrestrische Physisk, Giessenbachstrasse 1, 85748 Garching, Germany\and
		      Leiden Observatory, Leiden University, PO Box, 2300 RA Leiden, The Netherlands\and
		      INAF, Osservatorio Astronomico di Cagliari, via della Scienza 5, 09047 Selargius, Italy\and
		      University of Vienna, Department of Astronomy, T\"{u}rkenschanzstrasse 17, 1180 Vienna, Austria
		     }

	   \date{}

	 
	  \abstract
{The mid-IR detection rate of water lines in disks around Herbig stars disks is about 5\%, while it is around 50\% for disks around TTauri stars. The reason for this is still unclear.}
{In this study, we want to find an explanation for the different detection rates between low mass and high mass pre-main-sequence stars (PMSs) in the mid-IR regime.}   
{We run disk models with stellar parameters adjusted to spectral types B9 through M2, using the radiation thermo-chemical disk modeling code ProDiMo. We produce convolved spectra at the resolution of Spitzer IRS, JWST MIRI and 
VLT VISIR spectrographs. We apply random noise derived from typical Spitzer spectra for a direct comparison with observations.} 
{The strength of the mid-IR water lines correlates directly with the luminosity of the central star. We explored a small parameter space around a standard disk model, considering dust-to-gas mass ratio, disk gas mass,
mixing coefficient for dust settling, flaring index, dust maximum size and size power law distribution index. 
The models show that it is possible to suppress the water emission; however, current observations are not sensitive enough to detect mid-IR lines in disks for most of the explored parameters.
The presence of noise in the spectra, combined with the high continuum flux (noise level is proportional to the continuum flux), is the most likely explanation for the non detections towards Herbig stars.}
{Mid-IR spectra with resolution higher than 20000 are needed to investigate water in protoplanetary disks. Intrinsic differences in disk structure, e.g. inner gaps, gas-to-dust ratio, dust size and 
distribution, and inner disk scale height, between Herbig and TTauri star disks are able to explain a lower water detection rate in disks around Herbig stars.}

{}

{}


	   \maketitle
	%

\section{Introduction}\vspace{5mm}

Water has been detected successfully towards many T~Tauri disks in the IR \citep{salyk,carr,pontoppidan1,pontoppidan2,bergin2,hogerheijde,riviere-marichalar}. Contrary, disks around Herbig stars do not show warm water lines
both in mid-IR \citep[15.17, 17.22, 29.85~$\mu$m blends from][]{pontoppidan1}, and in the near-IR \citep[2.9345~$\mu$m ro-vibrational lines from][]{fedele1}. Far-IR water lines have been detected towards the Herbig star HD~163296
\citep{meeus2,fedele}; tentative detections are reported towards other targets such as HD 31648 (63~$\mu$m), HD~97048 and HD 100546, but the blend with the CH$^+$ at 90 $\mu$m and 179.5~$\mu$m prevent a firm claim for the 
detection of water at this wavelength \citep{meeus2}. Finally six successful detections of the OH molecule have been reported from the Keck II NIRSPEC instrument towards AB~Aur and MWC~758  \citep{mandell}, and from CRIRES
towards V380~Ori, HD~250550, HD~259431, HD~85567 \citep{fedele1}.

Several possible explanations have been considered for the non detections of water and other molecular features towards Herbig stars: veiling due to stronger continuum \citep{pontoppidan2}, difference 
in dust content of the emitting layers \citep[due to a different dust-to-gas mass ratio or settling,][]{meijerink}, different efficiency in the heating of the surface layers, a difference in the chemistry between Herbig 
and T~Tauri disks, the transitional nature of the observed disks. The inner disk physical conditions seems to be independent from the mass of the central star \citep{salyk,mandell}, since the OH and water lines are 
consistent with the same gas temperature and location of the emitting regions of both T~Tauri and Herbig stars. We know that some disks around Herbig stars are transitional, and in some of these OH, CO-rovibrational and OI 
lines have been detected \citep[e.g. HD~100546][]{liskowsky}. However, in some cases the inner cavity is not totally gas depleted, and transitions as the CO-rovibrational band at 5~$\mu$m are still detectable 
\citep{salyk3}. 
The last hypothesis, the role of FUV radiation field in photodissociating water in the inner disk seems supported by NIRSPEC \& CRIRES observations of OH \citep[respectively,][]{mandell,fedele1}. Multi-epoch Spitzer
spectra of the variable star EX~Lup \citep[][]{banzatti1} support the idea that photochemistry must be responsible for organics depletion, enhanced OH column density and reduced water column density.

Modeling of Herbig stars performed with the radiation thermo-chemical code ProDiMo \citep{woitke1} suggests that disks around Herbig Ae stars are not dry, and are still able to emit far-IR lines. 
The presence of the gaseous reservoir in the outer disk is due to an interplay between photodissociation and photodesorption. Due to this interplay, water vapour is present even in disks around bright stars (as the Herbigs, that 
have stronger FUV flux), but the water ice reservoir is less extended with respect to the case of disks around low mass stars \citep{vandishoeck2}. The inner disk water vapour reservoir is radially more extended in disks around 
earlier type stars \citep[higher effective temperature, $T_\mathrm{eff}$,][]{du}.

In a previous work \citep{antonellini} we investigated the effects of different disk parameters on the chemistry and water spectroscopy in the mid- and far-IR. We identified key parameters affecting the continuum 
opacity and the gas content of the disk as drivers of the mid- and far-IR lines. The explanation for low detection rate of the far-IR water lines with Herschel/HIFI is likely sensitivity; the observations were often not 
deep enough. 

In this work, we focus on the difference between low and high mass PMSs. We searched for factors which can be responsible for the absence of mid-IR water lines in disks around Herbig stars, comparing the results with what 
we find for T~Tauri and intermediate mass PMS. We model here a different central star, inner disk properties and instrumental properties. In Section~\ref{2}, we describe the code and modeling approach. In Section~\ref{3}, we 
report our results. In Section~\ref{4}, we discuss the results and finally in Section~\ref{5}, we present our conclusions.
   

\section{Modeling}\vspace{5mm}
\label{2}

We perform modeling of protoplanetary disks using the radiation thermo-chemical code ProDiMo (``Protoplanetary Disk Model'') including X-ray photoprocesses \citep{woitke,aresu}. The code computes self consistently the 
chemistry (steady state) with the heating/cooling balance. Heating processes implemented include PAH/photoelectric heating, C photoionization, H$_2$ photodissociation, formation heating and collisional de-excitation, cosmic rays
heating, viscous heating, pumping by OH photodissocation, H$_2$ pumping by formation on dust grains, and chemical heating and several others. Cooling processes occur through more than 10000 atomic/molecular emission lines. Finally, 
gas and dust thermalization is also taken into account.
Other features include a computation of the photo-rates using detailed cross-sections and an extension of the collisional partners for water \citep{kamp4}, a soft-edge description of the outer disk, the PAH ionization 
balance and heating/cooling, UV fluorescence pumping, a parametric description of settling \citep{woitke1}.

We computed here the detailed statistical equilibrium of a subset of 500 ro-vibrational levels, from the complete set \citep{tennyson,barber}, which includes up to 411 o-H$_2$O and 413 p-H$_2$O ro-vibrational levels (rotational 
levels up to $J$~=~25; projections up to $K_\mathrm{a,c}$~=~18; vibrational levels up to $v_1$~=~2, $v_2$~=~2, $v_3$~=~1). The study here focuses on the 12 $\mu$m series of lines observed from the ground by \citet{pontoppidan2}
and described also in our previous work \citep{antonellini}. In addition, we consider the blends centered at 15.17~$\mu$m observed by Spitzer IRS and reported in \citet{pontoppidan1}. 
Our line fluxes are derived from level populations computed through vertical escape probability. This approach has been used also in our previous work \citep{antonellini} and it provides line fluxes trustable within a 
factor two from a more computationally expensive detailed radiative transfer. The transitions we consider in this work are from very warm gas layers, and have $E_\mathrm{up}$ larger than 500~K. The transitions contributing to the 
12.407~$\mu$m blend have $E_\mathrm{up}~>~3000$~K, while the lines contributing to the 15.17~$\mu$m blend have $E_\mathrm{up}~>~2700$~K.

\subsection{Central star}

For the basic model, we use the same standard disk as in our previous work on TTauri stars (properties are described in Table~\ref{global}). We then consider different central star spectra corresponding to real observed 
PMS targets, including the non photospheric contributions due to accretion\footnote{In this work as in the previous, we did not treat the presence of UV excess consistently with the presence of viscous heating. Our disks are 
passive. FUV excess is considered as an extra UV radiation field emitted isotropically from the central star.} 
and X-rays. The spectra are built constraining the photometric observations in the optical and the UV by fitting $T_\mathrm{eff}$, $R_\mathrm{V}$, $p_\mathrm{UV}$, $f_\mathrm{UV}$, using literature values for the distance and 
$M_\mathrm{star}$ (Woitke et al. submitted).
We use the Kurucz and Phoenix stellar atmosphere libraries \citep{husser,murphy}, and the high energy spectra merging IUE, HST and FUSE observations for the FUV, and XMM Newton, Chandra for the X-rays after a post-processing
described in Dionatos et al. in prep. FUV data for the stars we considered are in Appendix~\ref{app2}.

The 10 different central stars we consider are based on observed available data (Table~\ref{stars}). The sample includes spectral types from M2 to B9 in order to probe the effects of a different photospheric temperature and 
luminosity, but also different non photospheric contributions. The input spectra used for the code are the result of the merging of three spectral regions: photospheric (optical, near-IR, photospheric UV), UV excess, X-rays.
Every of these contributions has its own luminosity: $L_\mathrm{star}$ is the photospheric luminosity, $L_\mathrm{UV}$ is the UV excess (from 3300~\AA~up to 1150~\AA, variable from source to source), $L_\mathrm{X}$ is the X-ray
luminosity (from 5~\AA~up to 35~\AA, variable from source to source). We check the consistency of the fitted stellar parameters with the Siess' track isochrones \citep{siess}\footnote{http://www.astro.ulb.ac.be/~siess/pmwiki/pmwiki.php/WWWTools\\/Isochrones}, 
taking the published age of the star and the mass, and verifying the consistency with the expected $T_\mathrm{eff}$, $L_\mathrm{star}$ and $R_\mathrm{star}$.

\begin{table*}
\caption{Central star properties} 
\centering
\resizebox{\textwidth}{!}{
\begin{tabular}{|c|c|c|c|c|c|c|c|c|c|}
\hline\hline
Star & Sp.Type$^{(*)}$ & $M_\mathrm{star}$ [M$_{\odot}$] & $d$ [pc] & $A_\mathrm{V}$ [mag] & $R_\mathrm{star}$ [R$_{\odot}$] & $T_\mathrm{eff}$ [K] (fit) & $L_\mathrm{star}$ [L$_{\odot}$] (fit) & $L_\mathrm{FUV}/L_\mathrm{star}$ & $L_\mathrm{X-ray}/L_\mathrm{star}$\\ \hline
\small{HD 100546} & \small{B9V} & 2.50$^{(1)}$ & 103$\pm6^{(2,26)}$ & 0.36$^{(20)}$ & 1.9$^\mathrm{(Siess)}$ & 11412 & 42.3 & 0.226 & 9.779$\cdot$10$^{-7}$\\
\small{HD 97048} & \small{A0} & 2.50$\pm0.20^{(3,25,27)}$ & 158$^{(25)}$ & 1.24$^{(1)}$ & 2.12$^{(25)}$ & 10000 & 33.5 & 0.151 & 4.345$\cdot$10$^{-7}$\\
\small{HD 163296} & \small{A4 (A1Ve)} & 2.47$^{(5)}$ & 122.0$^{(4,25,26,27)}$ & 0.5$^{(19)}$ & 2.3$\pm0.1^{(19)}$ & 8907 & 38.3 & 0.088 & 5.345$\cdot$10$^{-7}$\\
\small{HD 142666} & \small{A8V} & 1.6$^{(6)}$ & 145$\pm43^{(27)}$ & 0.8$^{(6)}$ & 1.483$^{(21)}$ & 7297 & 6.4 & 0.030 & 2.114$\cdot$10$^{-6}$\\
\small{HD 135344B} & \small{F5V (F8V)} & 1.70$_{-0.1}^{+0.2 (8)}$ & 140.0$^{(25,26,27)}$ & 0.4$^{(17)}$ & 2.2$^{(25)}$ & 6950 & 8.1 & 0.007 & 2.871$\cdot$10$^{-5}$\\
\small{RY Tau} & \small{F8V (F8Ve-K1IV-Ve)} & 2.24$\pm0.07^{(24)}$ & 140.0$^{(9)}$ & 2.2$^{(9)}$ & 2.9$^{(9)}$ & 5496 & 10.9 & 0.067 & 5$\cdot$10$^{-4}$\\
\small{RY Lup} & \small{G0V} & 1.38$^{(11)}$ & 120$\pm35^{(28)}$ & 2.48 & 1.67$^{(11)}$ & 5200 & 2.6 & 0.221 & 9.845$\cdot$10$^{4}$\\
\small{DN Tau} & \small{K6V} & 0.65$\pm0.05^{(12)}$ & 140.0$^{(14)}$ & $\simeq0.5^{(12)}$ & 1.9$\pm0.2^{(12)}$ & 3904 & 0.8 & 0.002 & 6.946$\cdot$10$^{-4}$\\
\small{CY Tau} & \small{M2V (M1.5)} & 0.48$\pm0.05^{(24)}$ & 140.0$^{(14)}$ & 0.1$^{(16,24)}$ & 1.4$^{(16)}$ & 3628 & 0.4 & 0.003 & 7.482$\cdot$10$^{-5}$\\
\small{DO Tau} & \small{M6 (GV:e)} & 0.56$\pm0.05^{(24)}$ & 140.0$^{(14,16)}$ &2.64$^{(24)}$ & 2.4$^{(16)}$ & 3777 & 1.3 & 0.090 & 1.07$\cdot$10$^{-4}$\\ \hline
\end{tabular}
}
\tablefoot{For all stars, we assumed $R_\mathrm{V} = 3.1$. The radius for HD100546 has been retrieved from the Siess evolutionary track \citep{siess}.}
\tablebib{(*) http://simbad.u-strasbg.fr/simbad/sim-fid; (1) \citet{vandenancker}, (2) \citet{perryman}, (3) \citet{vanleeuwen}, (5) \citet{tilling}, (6) \citet{garcia}, 
(8) \citet{muller}, (9) \citet{lopez-martinez}, (11) \citet{manset}, (12) \citet{donati}, (14) \citet{luhman}, (16) \citet{hartigan}, (17) \citet{carmona}, 
(19) \citet{ellerbroek}, (20) \citet{ardila}, (24) \citet{bertout1}, (25) \citet{maaskant}, (26) \citet{mcjunkin}, (27) \citet{vanboekel}, 
(28) \citet{fedele2}}
\label{stars}
\end{table*}

\subsection{Parameter space exploration of a disk around a typical Herbig star}

We build also a small parameter series, exploring a subset of disk properties listed in Table~\ref{global}: scale height radial power law index $\beta$, dust-to-gas mass ratio, disk gas mass, scale height of the disk at the 
inner disk radius, inner disk radius, mixing coefficient for the dust settling \citep[accordingly to the settling prescription from][]{dubrulle}, dust size distribution power law index, dust grains maximum size. Each model 
contains all the quantities fixed and equal to our standard model ones, except the ones indicated in Table~\ref{global}. A more detailed description of the previous disk properties can be found in \citet{antonellini}.
This kind of exploration allows us to disentangle effects of single parameters on disks irradiated by a central Herbig star. In this small parameter series, the disk is irradiated by a typical Herbig star (2.2~M$_{\odot}$, 
32 L$_{\odot}$, 8600 K). The disk considered is the same as the standard disk model of our previous work \citep{antonellini}, with the exception of the inner radius, which is computed consistently with the 
sublimation temperature of the dust ($R_\mathrm{in}~=~0.365$~au; Table \ref{global}).

\begin{table*}
\caption{Herbig disk grid}
\centering
\begin{tabular}{p{5.0cm}p{3.3cm}p{7.5cm}} 
\hline\hline
\multicolumn{3}{c}{Standard disk model parameters}\\
\hline
Parameter & Symbol & Value\\ \hline
Grid points & $N_\mathrm{xx}$/$N_\mathrm{zz}$ & 70 $\times$ 70\\
Outer radius & $R_\mathrm{out}$ [au] & 300\\ 
Surface density power law index & $\epsilon$ & 1.0\\ 
Minimum dust size & a$_\mathrm{min}$ [$\mu m$] & 0.05\\ 
Tapering-off radius & $R_\mathrm{taper}$ [au] & 200\\ 
Chemical Heating & - & 0.2\\ 
Settling description & - & Dubrulle \\ 
Cosmic ray ionization rate & $\zeta_\mathrm{CRs}$ [s$^{-1}$] &
1.7$\times10^{-17}$\\
distance & $d$ [pc] & 140\\
Turbulence viscosity coefficient & $\alpha_\mathrm{vis}$ [-] & 0\\
Dust properties & - & astrosilicates$^{(1)}$ \\
\hline\hline
\multicolumn{3}{c}{Disk parameters varied in the series of models irradiated by a typical Herbig star$^{(2)}$}\\
\hline
Parameter & Symbol & values\\ \hline
Flaring power law index & $\beta$ & -0.8, -0.5, -0.1, 0.0, 0.8, 0.85, 0.9, 0.95, 1.0, 1.05, 1.1, $\bf{1.13}$, 1.15, 1.2, 1.25\\ 
Dust-to-gas mass ratio & $d/g$ & 0.001, $\bf{0.01}$, 0.1, 1, 10, 100\\ 
Disk gas mass & $M_\mathrm{gas}$ [M$_{\odot}$] & 10$^{-5}$, 10$^{-4}$, 0.001, $\bf{0.01}$, 0.05, 0.1 \\ 
Inner disk scale height (at $R = R_\mathrm{in}$) & $H_{0}$ [au] & 1.8(-3), 4.4(-3), 9.0(-3), $\bf{1.76(-3)}$, 3$\times$10$^{-2}$, 0.1\\ 
Inner radius & $R_\mathrm{in}$ [au] & 0.1, $\bf{0.365}$, 0.5, 1.0, 5.0, 10.0, 15.0, 20.0, 25.0, 30.0\\ 
Mixing coefficient & $\alpha_\mathrm{set}$ & 10$^{-5}$, 10$^{-4}$, 0.001, 0.01, $\bf{0.05}$, 0.1\\ 
Dust power law index & $a_\mathrm{pow}$ & 2.0, 2.5, 3.0, $\bf{3.5}$, 4.0, 4.5 \\
Dust maximum size & $a_\mathrm{max}$ [$\mu$m] & 250, 400, 500, 700, $\bf{1000}$, 2000, 5000, 10$^4$, 10$^5$\\
Elemental abundance of O & $\Delta Y_\mathrm{O} = Y_\mathrm{O,i}-Y_\mathrm{O,standard}$ & -0.5, -0.4, -0.3, -0.2, -0.1, $\bf{0.0}$, 0.1, 0.2, 0.3, 0.4, 0.5\\ \hline
\end{tabular}
\tablefoot{Bold numbers refer to standard model values. In parenthesis are shown the orders of magnitude for the values of the models with different scale height.
(1) \citet{draine2}; (2) $M_{star}~=~2.2$~M$_{\odot}$, $L_{star}~=~32$~L$_{\odot}$, $T_\mathrm{eff}~=~$8600~K}
\label{global} 
\end{table*}


\section{Results}\vspace{5mm}
\label{3}

\subsection{Series 1: {\bf{D}}ifferent central star}

\begin{figure}[htpl!]
\centering
\includegraphics[width=0.48\textwidth]{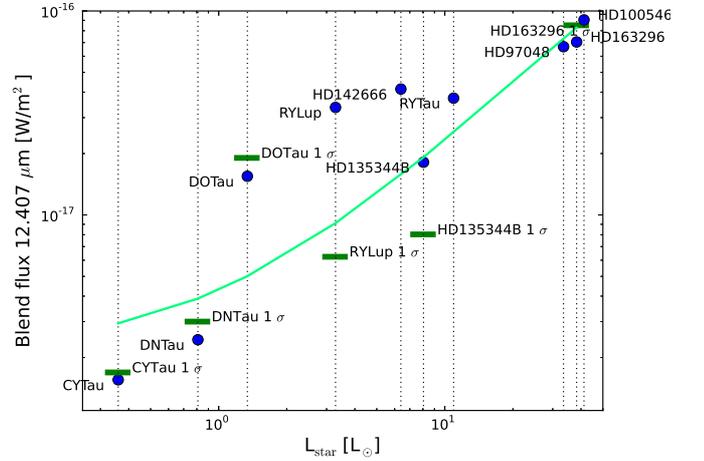}
\caption{Line fluxes and luminosities from the central star versus effective temperature for Spitzer 12.407~$\mu$m blend. The blue dots are the line fluxes of the models for the different central stars. 
The stars mark, the luminosities in the different wavelength bands. Detection limits for Spitzer IRS observations towards the same central star disks are plotted as thick green dashes (see Appendix~\ref{app1}). Light green 
curve is a fits to the sensitivity limits. The central stars are labeled with the name reported in column 1 of Table~\ref{global}.}
\label{excesses}
\end{figure}

\begin{figure}[htpl!]
\centering
\includegraphics[width=0.48\textwidth]{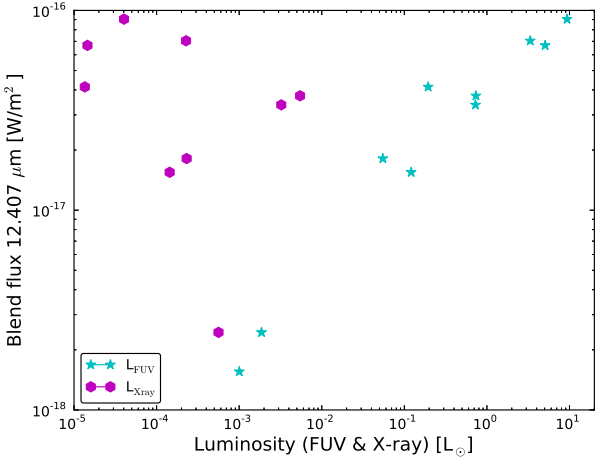}
\caption{Line fluxes for Spitzer 12.407 $\mu$m blend versus $L_\mathrm{FUV}$ (stars cyan) and $L_\mathrm{X}$ (magenta exagons).}
\label{bright}
\end{figure}

The general trend is shown in Fig.~\ref{excesses}, where the flux of the 12.407~$\mu$m water blend is plotted for our models with different central star. The mid-IR transitions get stronger with increasing photospheric 
temperature, although the progression is not linear. This can be explained by a combination of different stellar spectrum properties (e.g. hardness of the X-ray component, color of the FUV excess), that can affect the 
heating/cooling and or water abundance in the inner reservoir in a complex manner. In a previous study, \citet{aresu1} explored the effect of X-rays on far-IR water lines, finding that the 179~$\mu$m flux grows 
with $L_\mathrm{X}$ independently from the UV. The situation becomes more complex analyzing transitions produced in the inner disk. There is an $L_\mathrm{UV}$ threshold above which X-ray affects the far-IR 
water line fluxes. In our previous work \citep{antonellini}, we found that mid-IR transitions originate from regions in which water is produced by thermally activated channels and depleted by photodissocation. 
The spectroscopy in this region is affected by FUV radiation.
Overall, the line flux of the mid-IR lines correlates with the stellar bolometric luminosity (Fig.~\ref{bright}), and it is correlated with the FUV and uncorrelated with X-ray luminosities (blue and magenta symbols in the 
same plot). However the sample of stellar spectra studied is too small to go into a deeper analysis. 

\begin{figure}[htpl!]
\centering
\includegraphics[width=0.48\textwidth]{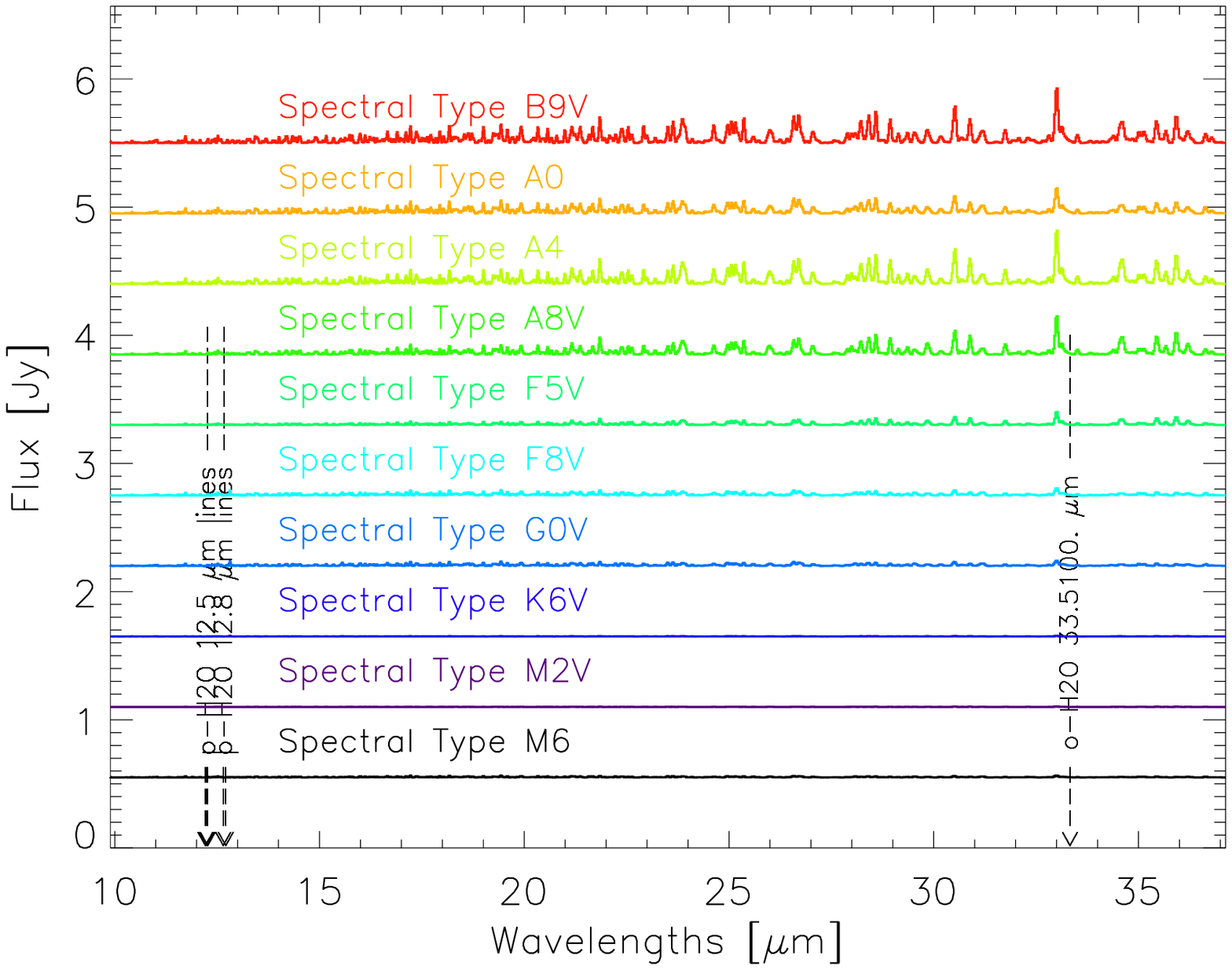}
\includegraphics[width=0.48\textwidth]{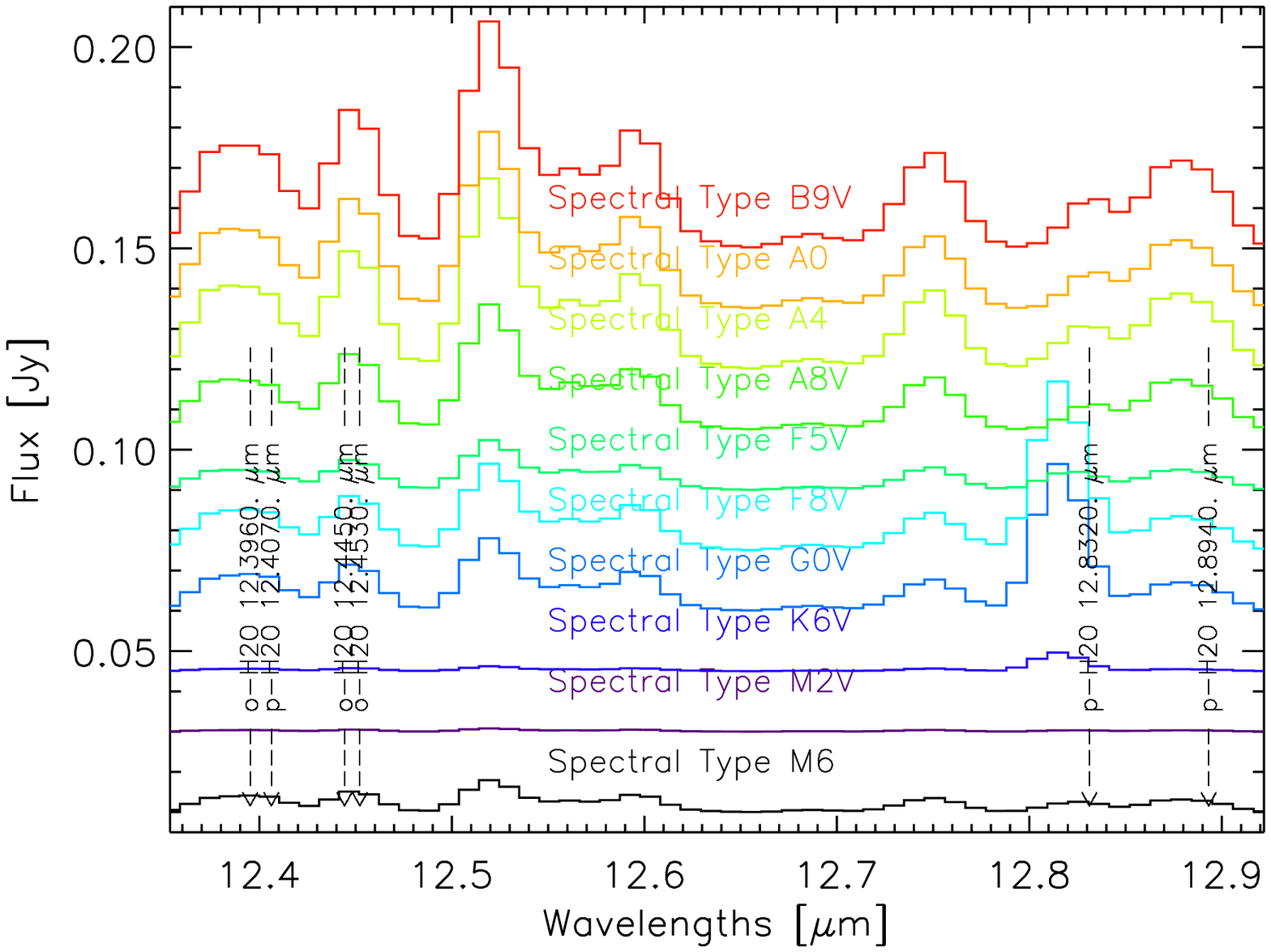}
\includegraphics[width=0.48\textwidth]{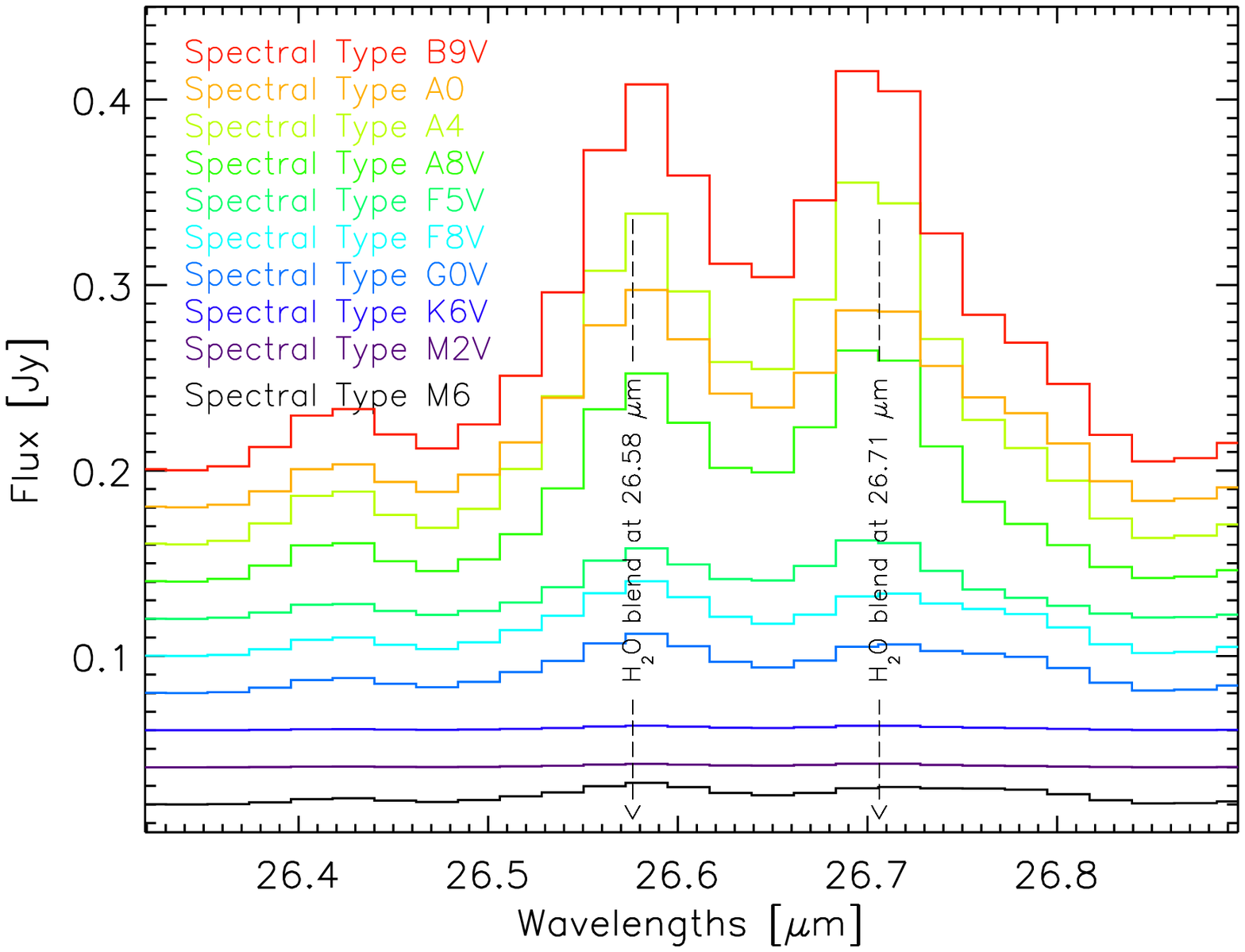}
\caption{Theoretical Spitzer SH/LH spectra (convolved at $R$~=~600 and rebinned) for models with different central stars; full continuum subtracted spectrum, artificially shifted for all the models (top), zoom on the 12~$\mu$m 
region (middle), zoom on the 26.5~$\mu$m region (bottom).}
\label{Spitzer-Teff}
\end{figure}

The theoretical Spitzer spectra (Fig.~\ref{Spitzer-Teff}) predict that the earliest spectral types have a richer mid-IR spectrum. A zoom into the 12~$\mu$m region shows that the trend is not only dictated by the spectral 
type: DN Tau (K6V) and CY Tau (M2V) are the only two objects with $L_\mathrm{star}~<~$1~L$_{\odot}$ (and FUV less than 5\% of the bolometric), and they show weaker mid-IR water lines. For example the M2V spectral type star 
($L_\mathrm{star}~=~$0.4~L$_{\odot}$) 12.407~$\mu$m flux is a factor 17 weaker than the M6 case ($L_\mathrm{star}~=$~1.3~L$_{\odot}$). This suggests again that the bolometric luminosity is the real driving force behind 
the mid-IR spectroscopy.
 
From Spitzer IRS observations towards the central stars we modeled (blue dots with thick green dashes in Fig.~\ref{excesses}), we find that the sensitivity anticorrelates with the stellar luminosity, indicating 
that observations around brighter stars are less sensitive.
The sensitivity limit for HD~163296 (Spectral type A4) in Spitzer spectra is very close to our standard disk model generated for the spectral type A4 (HD~163296 in Fig.~\ref{excesses}). The disk structure around this 
Herbig star has been robustly fitted by \citet{tilling} using multi-wavelength observations; it is very similar to our standard disk model. This may point to the fact that observations for this object were simply not deep enough
to detect the mid-IR water lines.


The modelled IRS spectra for different central stars seem to disagree with the observations towards a large sample of Herbig stars, whose spectra are featureless \citep{pontoppidan1}. The reason for the disagreement between 
observations and modeling can be twofold: either the structure of Herbig disks is different from TTauri disks, or instrumental effects affect differently the spectra of these two types of PMS disks. In the next two 
paragraphs, we test the first hypothesis comparing the result of modelling with observations, and then we apply instrumental characteristics to test the second hypothesis.

\subsection{Series 2: Rockline displacement effect}

\begin{figure}[htpl!]
\centering
\includegraphics[width=0.48\textwidth]{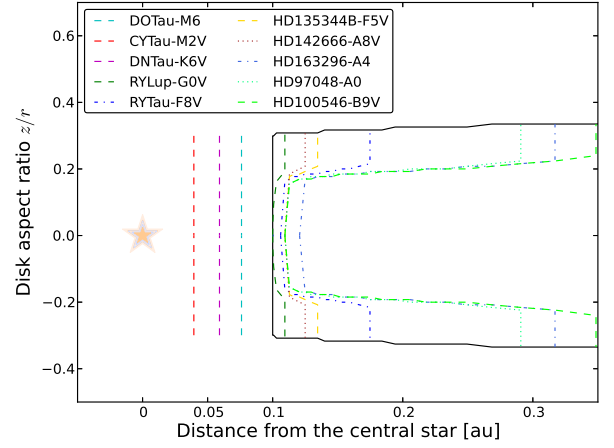}
\caption{Rockline position for the disks around different central stars. The thick black line indicates the disk contour at the gas density of 5$\cdot$10$^3$~cm$^{-3}$ for all our disk models, we consider it as delimitation of 
our disk, due to the fact that the erosion happens at more then 4 disk scale heights. 
The different color and line styles delimitate the contours of the disk with $T_\mathrm{dust}~=~1500$~K assumed as sublimation temperature of silicates, upward and inward of this contour the dust is sublimate, and the disk
is truncated. It describes the effective extension of the disk accordingly to the dust sublimation temperature. For late type stars disk temperatures are too low and the contour is replaced by a vertical line, that indicates the
distance from which the dust in thermal equilibrium should be photoevaporated. The position of the central star is indicated by the star symbol.}
\label{rock}
\end{figure}

The results from the standard disk irradiated by a different central star contain an inconsistency (since we wanted to isolate the effect of a changing central star from that of disk changes). The inner disk radius 
remains fixed at 0.1~au. Taking instead 1500~K as sublimation temperature for the dust (silicates) as adopted in \citet{min1}, Fig.~\ref{rock} shows how the disk inner radius is shifted outward for earlier type stars. For 
T~Tauri stars with $L_\mathrm{star}~<~1.5$~L$_{\odot}$, the inner radius is even truncated inward of 0.1~au.

\begin{figure}[htpl!]
\resizebox{\hsize}{!}{\includegraphics{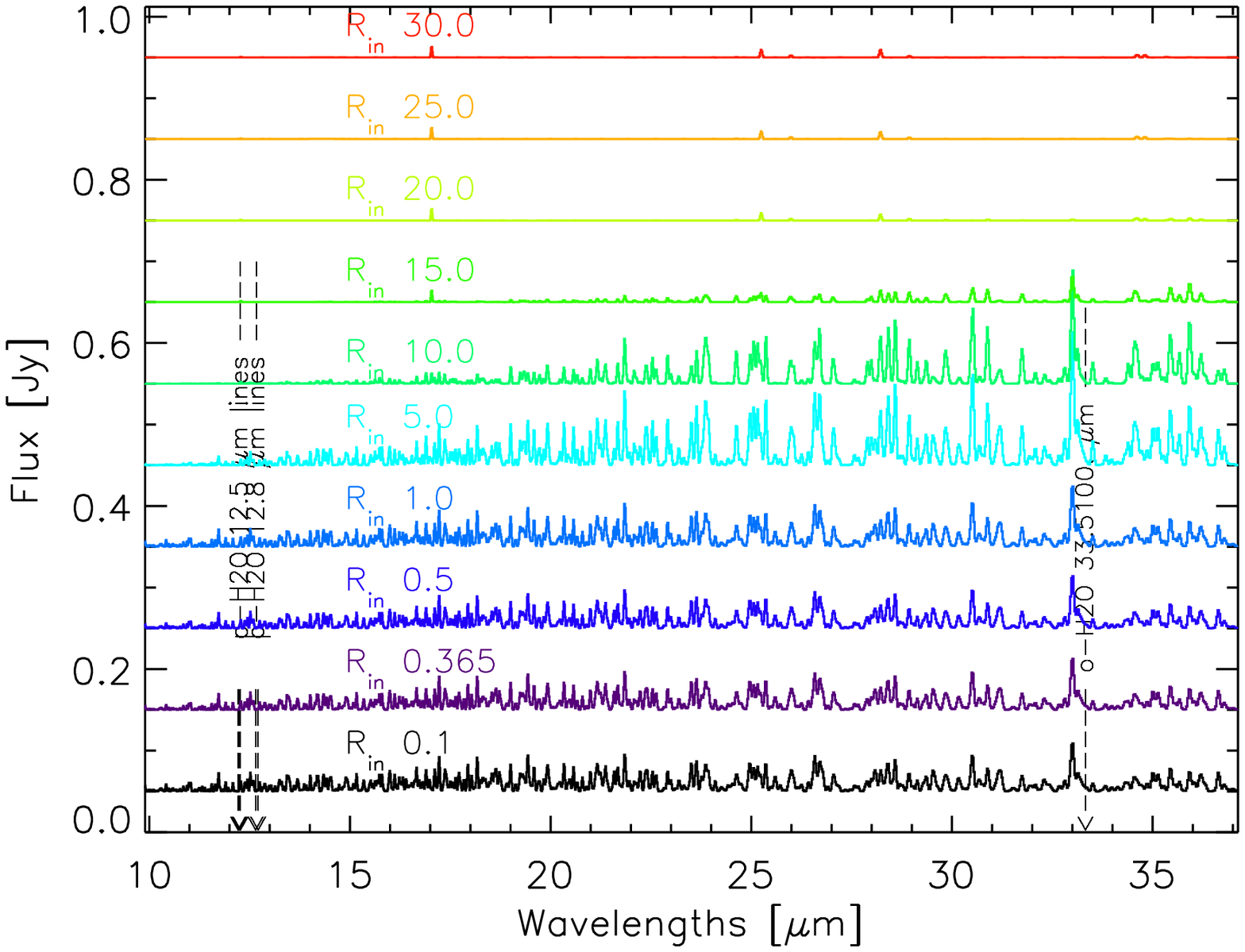}}
\resizebox{\hsize}{!}{\includegraphics{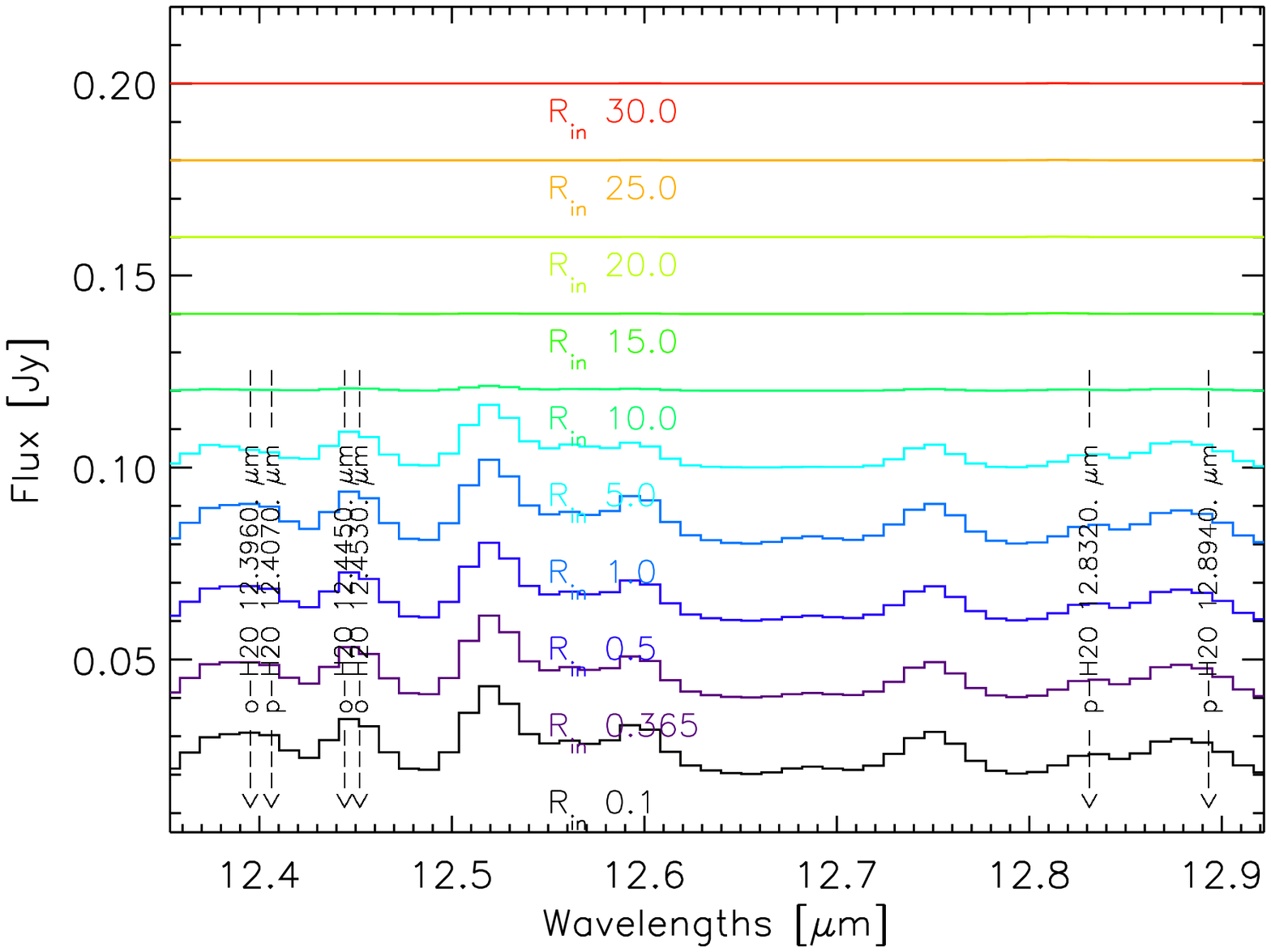}}
\resizebox{\hsize}{!}{\includegraphics{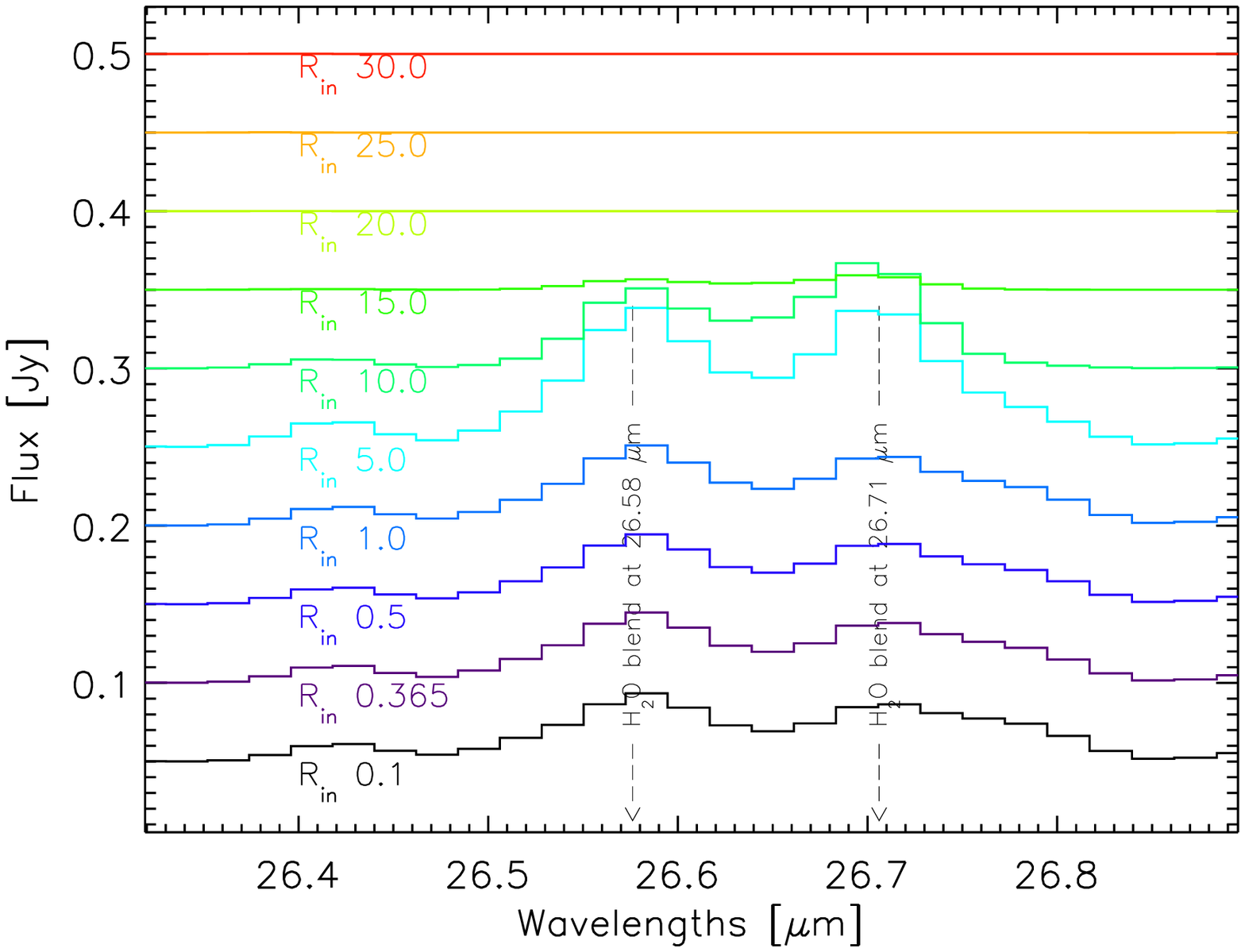}}
\caption{Continuum subtracted theoretical Spitzer spectra for $R_\mathrm{in}$ variation, artificially shifted, convolved at the resolution of $R~=~600$ and rebinned. Second plot is a zoom of the complete spectrum on the 
12~$\mu$m region. Third plot is a zoom on the 26.5~$\mu$m region.}
\label{Spitzer-Rin}
\end{figure}

In order to study the impact of a disk truncated by dust sublimation, we ran a series of models considering different disk inner radii and another one with different scale heights using always the same central star, a typical 
Herbig star (see Notes of Table.~\ref{global}). The larger inner disk radii produce a suppression of the whole mid-IR spectrum, in particular if the gap extends beyond 10~au (Fig.~\ref{Spitzer-Rin}). For inner radii less 
than 1.0~au, the Spitzer features become weaker again, due to a geometrical effect: the line is emitted further inward and the total emitting area becomes smaller.     
Increasing the scale height of the same disk, will strongly reduce the continuum opacity. As a consequence, the water ice reservoir is totally depleted and the water vapor reservoir is radially more extended (in particular the 
innermost one). The emitting water column density increases and mid-IR water lines become stronger. This is particularly true for disks with a scale height larger than 2$\times$10$^{-3}$~au at $R$~=~0.1 au.

\begin{figure}[htpl!]
\resizebox{\hsize}{!}{\includegraphics{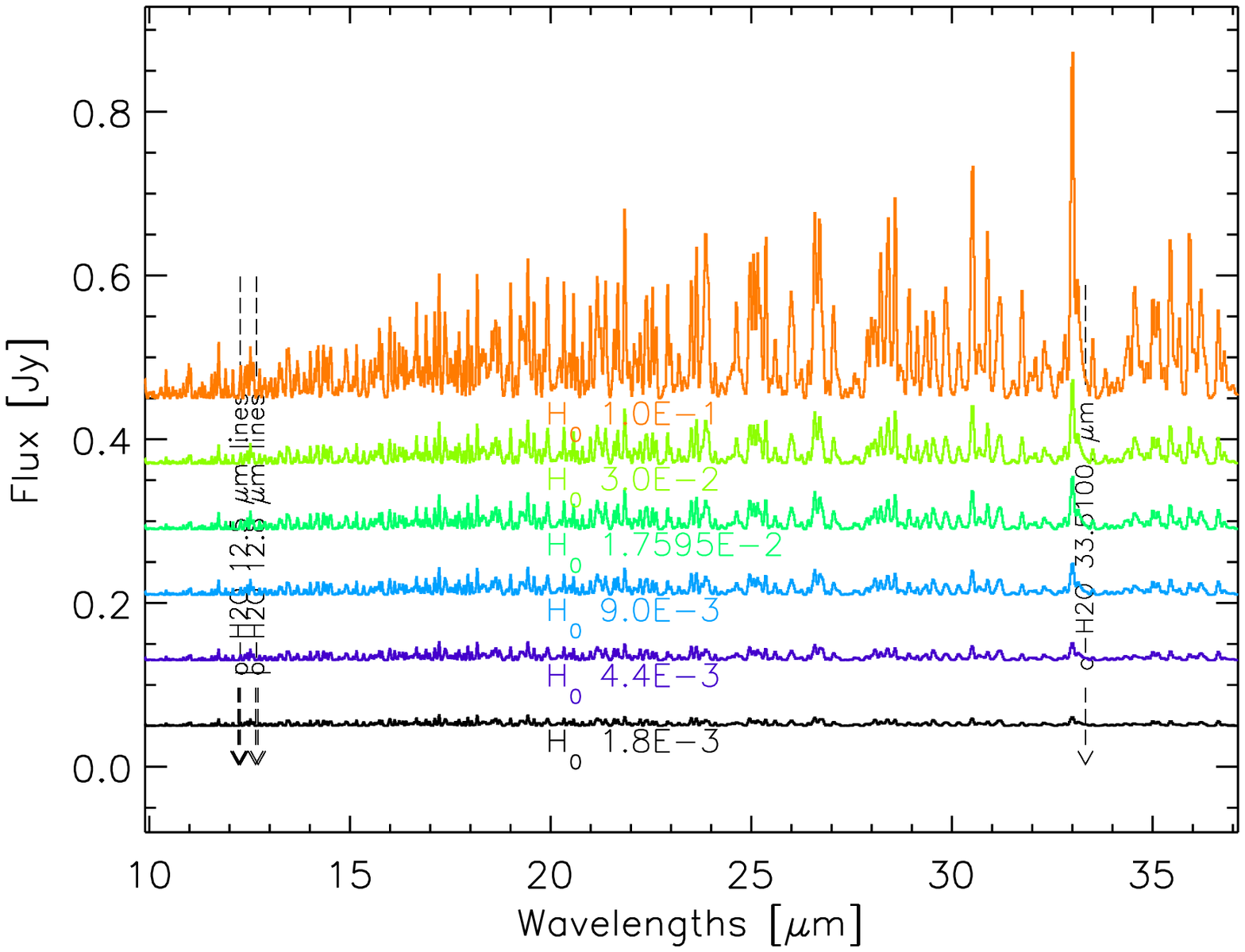}}
\resizebox{\hsize}{!}{\includegraphics{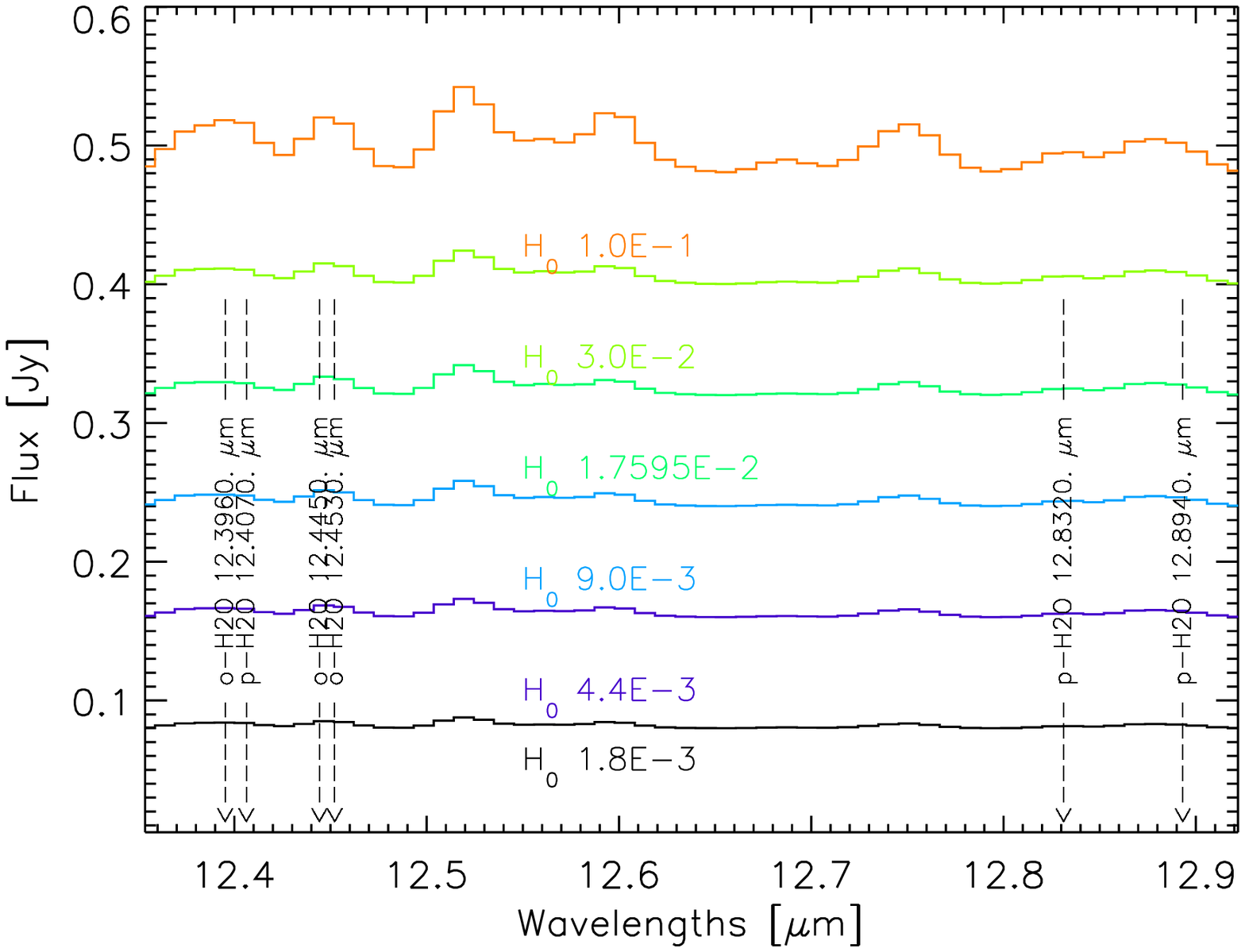}}
\resizebox{\hsize}{!}{\includegraphics{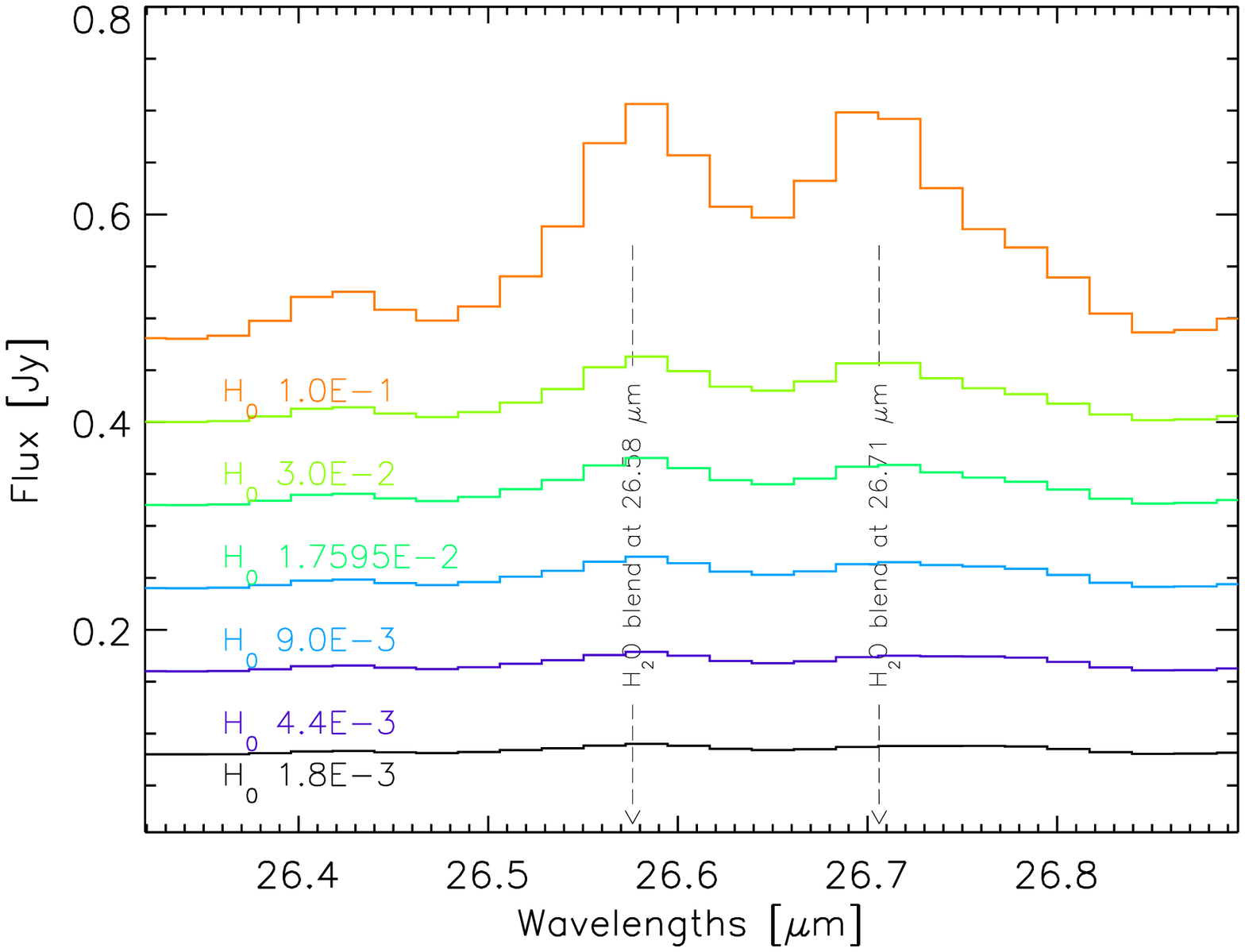}}
\caption{Continuum subtracted theoretical Spitzer spectra for $H_{0}$ variation, artificially shifted, convolved at the resolution of $R~=~600$ and rebinned. Second plot is a zoom of the complete spectrum on the 12~$\mu$m region.
Third plot is a zoom on the 26.5~$\mu$m region.}
\label{Spitzer-H0}
\end{figure}

The true rockline displacement is limited to tenth of au, and the lowering in scale height is about a factor 2 with respect to our standard disk for the brightest Herbig star we considered (Fig.~\ref{rock}).
In order to suppress the Spitzer water line emission, it would be necessary to open a gap larger than 5-10 au and contract the disk vertically by more than a factor 2. In particular, the 15.17~$\mu$m blend drops below the
typical sensitivity range we defined for Spitzer IRS observations of Herbig stars ({\bf{8$\times$10$^{-17}$-8$\times$10$^{-18}$}}~W/m$^2$) for $R_\mathrm{in}\!>\!20$~au and $H_0$ smaller than 9$\times$10$^{-3}$~au. This 
numerical test rules out that a simple difference in the disk structure is responsible for the water non-detections towards Herbig disks.

\subsection{Series 3: Extended parameter space exploration}

\begin{figure}[htpl!]
\includegraphics[width=0.48\textwidth]{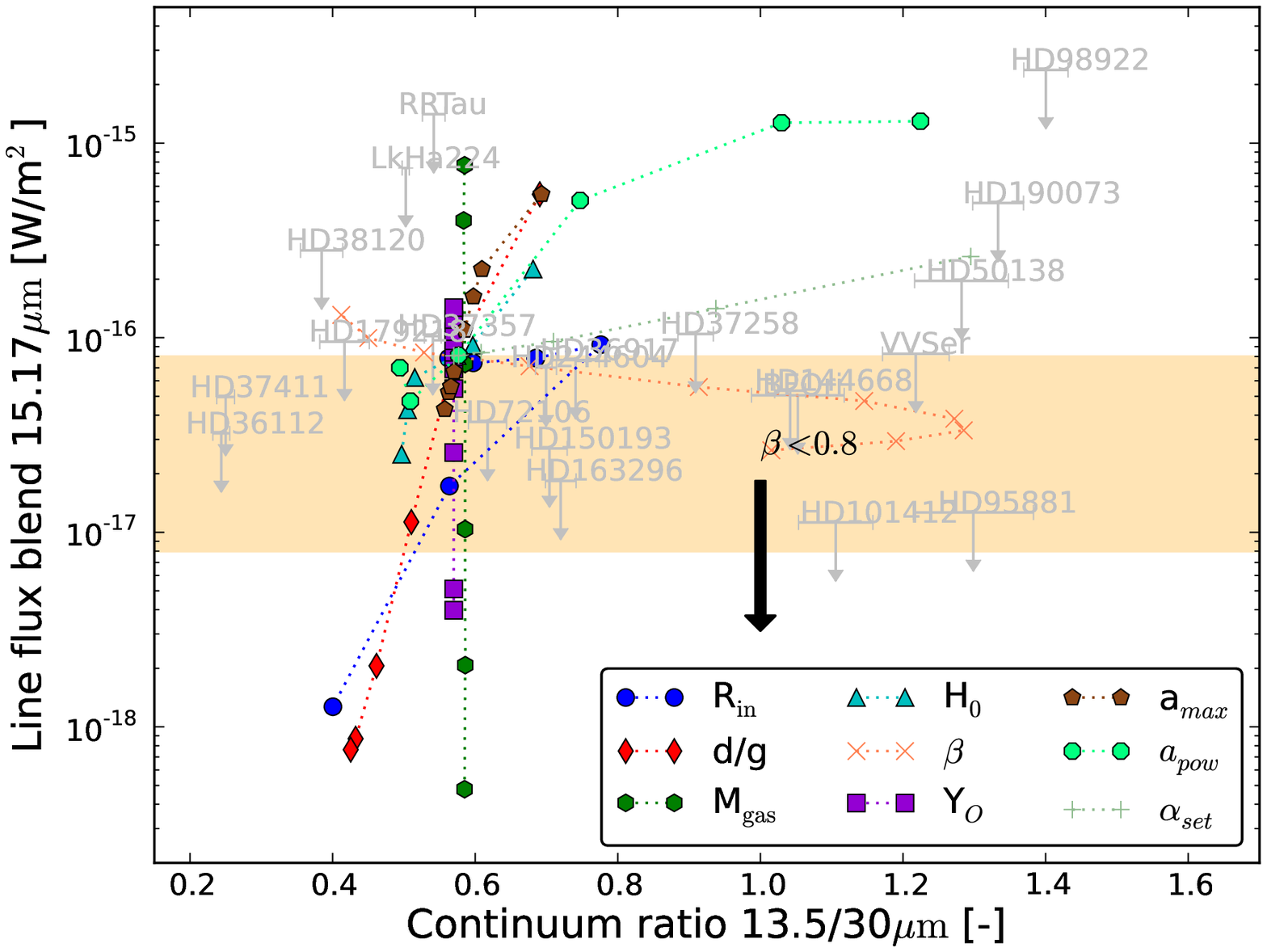}
\caption{Herbig parameter space for the water line blend at 15.17~$\mu$m. In the plot the data points (all upper limits, in grey) are from \citep{pontoppidan1}. All the fluxes are scaled to 140 pc. The shaded yellow area 
contains the range of sensitivity we retrieved from Spitzer spectra of Herbig stars HD~135344B and HD~100546. All the fluxes of the disks considered are scaled to 140~pc. The trend of disks flatter than $\beta~=~0.8$ is
shown with a black arrow, since too flat models will be self-shadowed, and the derived mid-IR color cannot be compared in a trustable manner with observations, in which the shadowing can be partial due to a more complex disk 
structure.}
\label{Space}
\end{figure}

For a direct comparison with Spitzer observations of Herbig disks, we expand the parameter space exploring also the dust-to-gas mass ratio, the flaring index, the gas mass, the C/O ratio, the settling \citep[through 
the $\alpha_\mathrm{set}$ turbulent coefficient from][]{dubrulle}, and the dust maximum size and power law distribution ($a_\mathrm{pow}$), around the typical central Herbig star (Sect.~\ref{2}). Fluxes extracted for mid-IR water 
blends at 15.17~$\mu$m \citep{pontoppidan1} are compared to the results from our series of models.
As mid-IR continuum tracer, we used the same ratio considered in our previous work \citep[13.5/30~$\mu$m,][]{antonellini}. Fig.~\ref{Space} shows the same plot as for the T~Tauri disk model series \citep[Fig.~13 top 
panel,][]{antonellini}. In our previous paper we showed that the transitions contributing to this blend behave in the same manner as our representative 12.407~$\mu$m line, and 
the same holds for the blends at 17 and 29~$\mu$m. Any conclusion on this blend will be representative of the entire mid-IR water spectrum.

The standard Herbig star disk, which is the base model of the parameter space exploration (Table~\ref{global}) is located in the central part of Fig.~\ref{Space}, where all the model series intersect.
We notice that a certain fraction of the targets and of our parameter space study is below the range of detection threshold deduced from the sensitivity limits reported in Fig.~\ref{excesses}. Part of the 
targets that would be detectable at 140~pc are too distant, and the exposure time length was not optimal for these objects. The fact that our typical Herbig model is close to the sensitivity range in Fig.~\ref{Space} 
suggests that instrumental factors may have played an important role in the chance of detecting water lines in Spitzer IRS observations.

\subsection{Instrumental effects}

The previous discussion is based on spectra unaffected by instrumental noise. Observations of water towards disks in the mid-IR wavelength range have been performed with the Spitzer Space Telescope and from the ground with the 
VLT \citep[e.g.\ ][]{pontoppidan2}. The future of near- and mid-IR water studies is JWST.
In order to take into account instrumental effects, we rebinned our oversampled modeled spectra and applied random noise effects (see Appendix~\ref{app}) to the convolved spectra \citep[Spitzer $R$~=~600, ][]{houck}.

\begin{figure}[htpl!]
\centering
\includegraphics[width=0.48\textwidth]{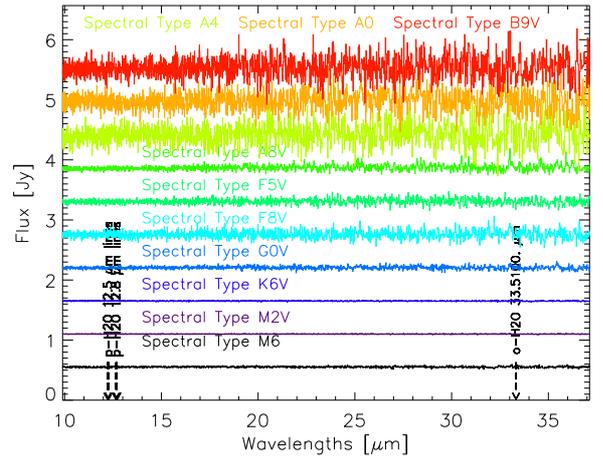}
\includegraphics[width=0.48\textwidth]{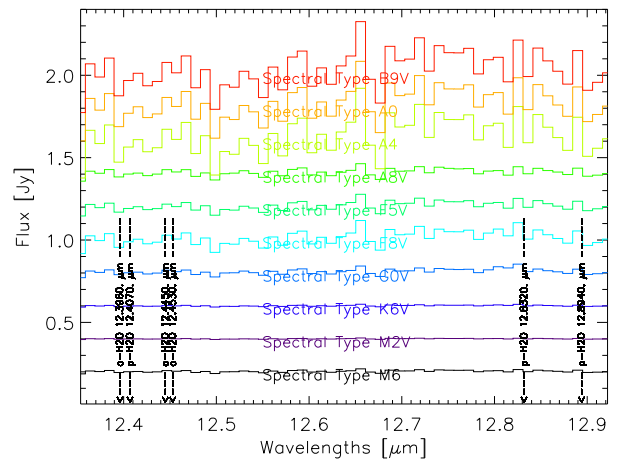}
\includegraphics[width=0.48\textwidth]{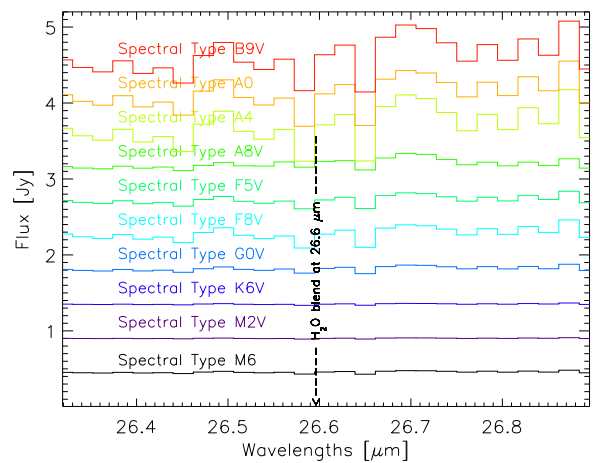}
\caption{Theoretical Spitzer SH/LH spectra ($R$~=~600) for models with different central stars rebinned and noisy; full continuum subtracted spectrum, artificially shifted for all the models (top), zoom on the 12~$\mu$m region 
(middle), zoom on the 26.5~$\mu$m region (bottom).}
\label{Sp-Teff-Noisy}
\end{figure} 

The rebinning procedure and the application of the random noise to the central stars models (Fig.~\ref{Sp-Teff-Noisy}) have the effect to suppress the water line fluxes in the mid-IR spectra. The effect is particularly dramatic 
for the earlier type stars, since they show a stronger continuum, and so a stronger noise. The possibility to use space or ground based instruments (in the mid-IR windows) with larger spectral resolution and sensitivity, would 
increase the detections of spectral features in the mid-IR towards both Herbig and T~Tauri stars.

\begin{figure*}[htpl!]
\centering
\includegraphics[width=0.48\textwidth]{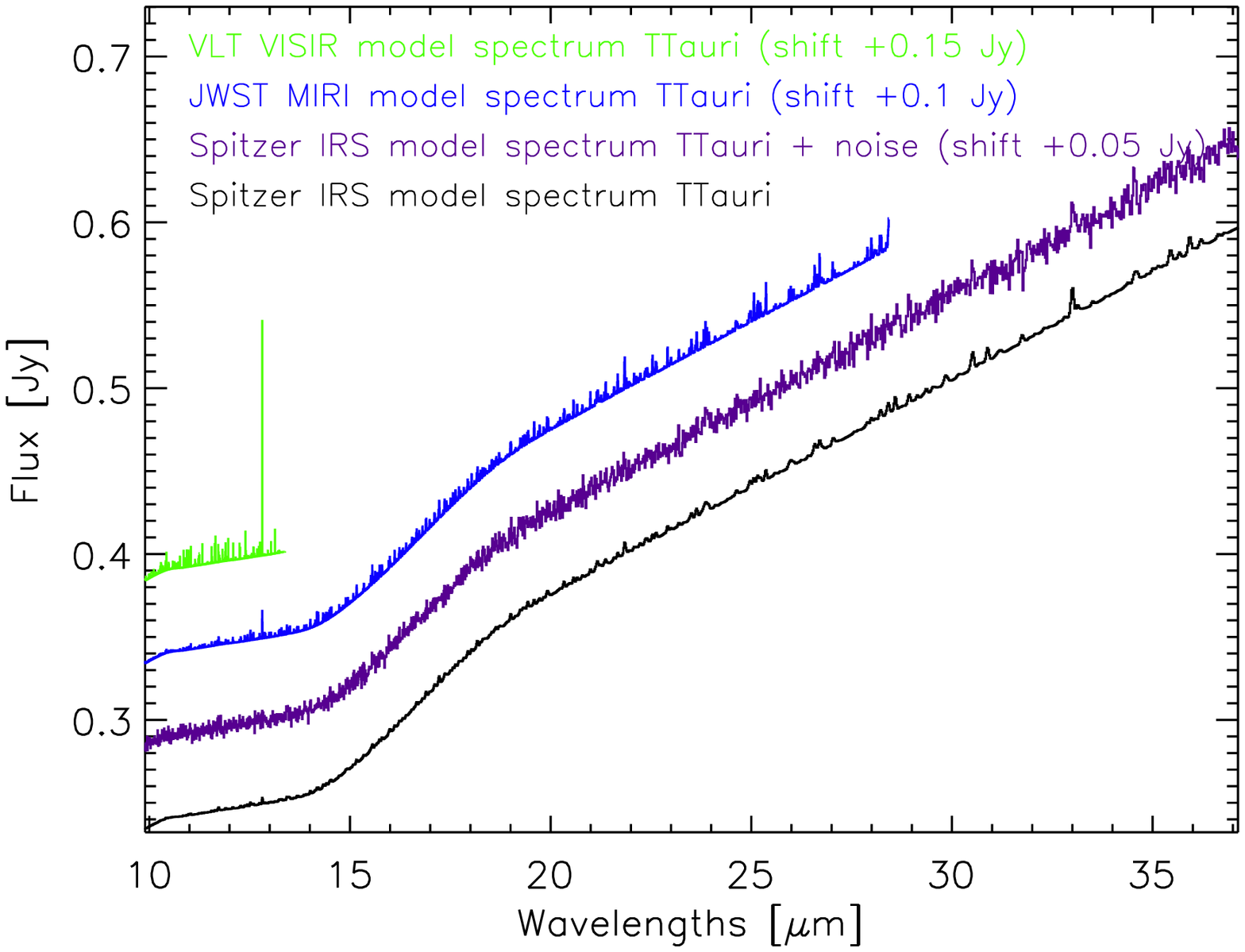}
\includegraphics[width=0.48\textwidth]{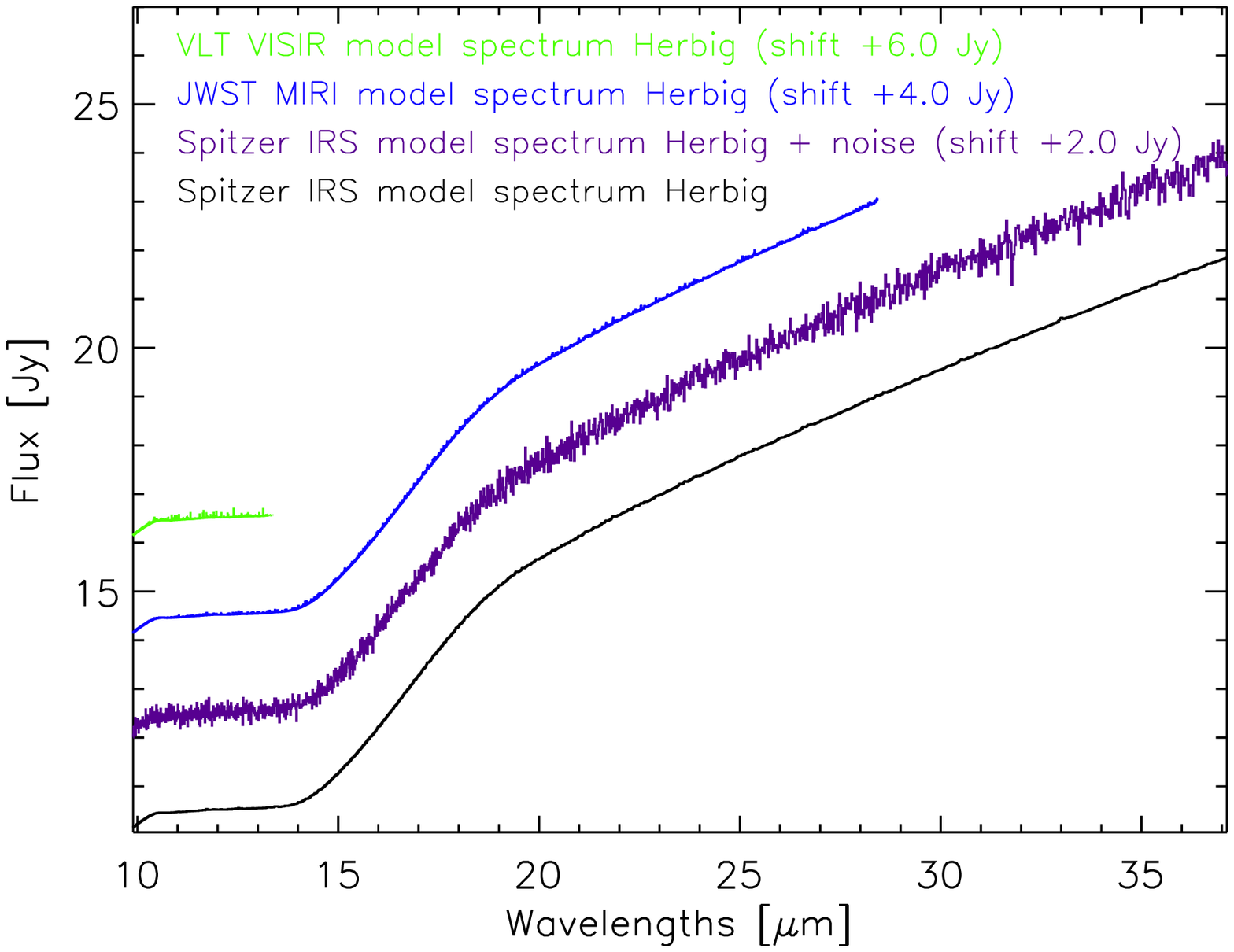}
\includegraphics[width=0.48\textwidth]{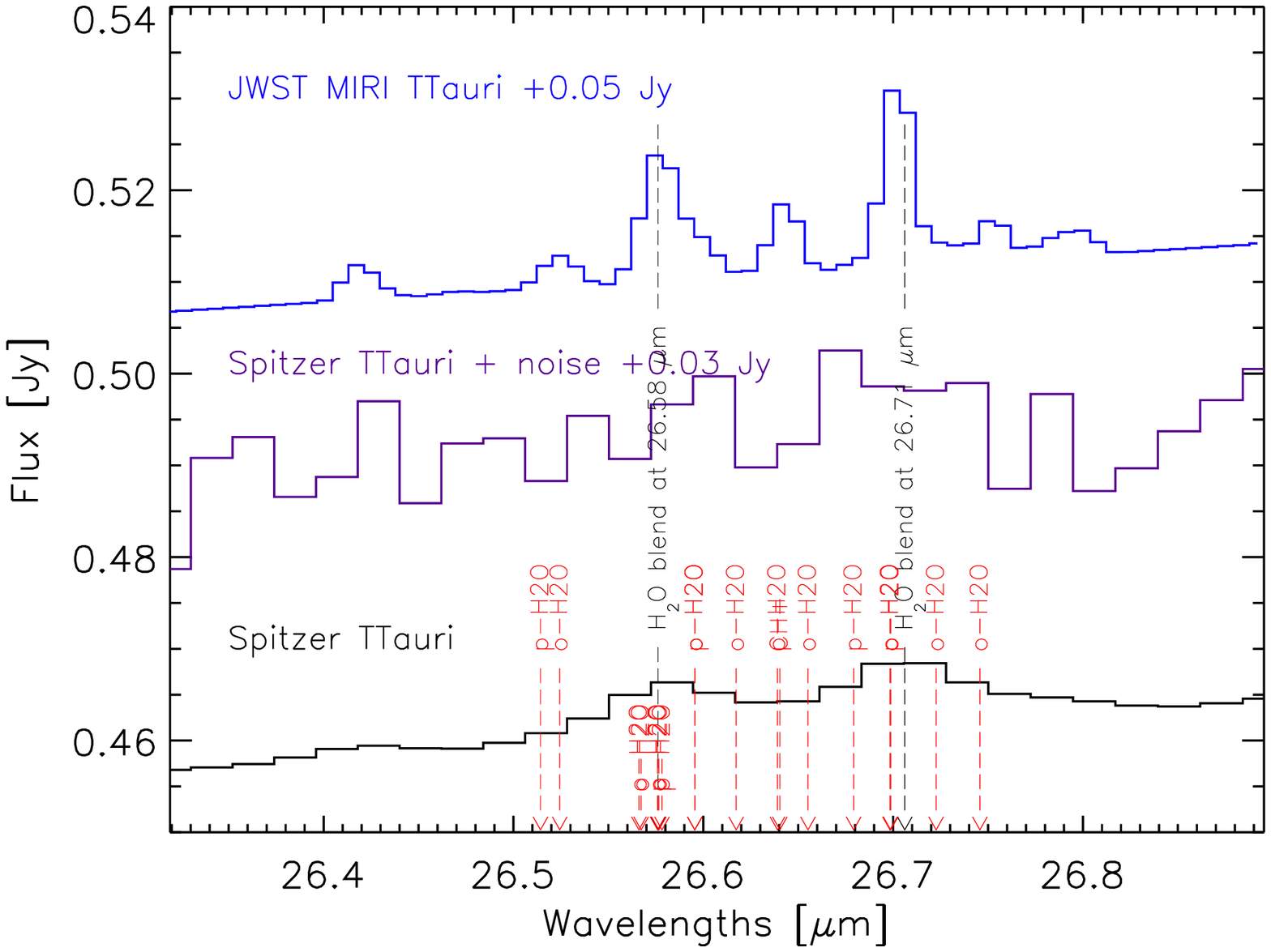}
\includegraphics[width=0.48\textwidth]{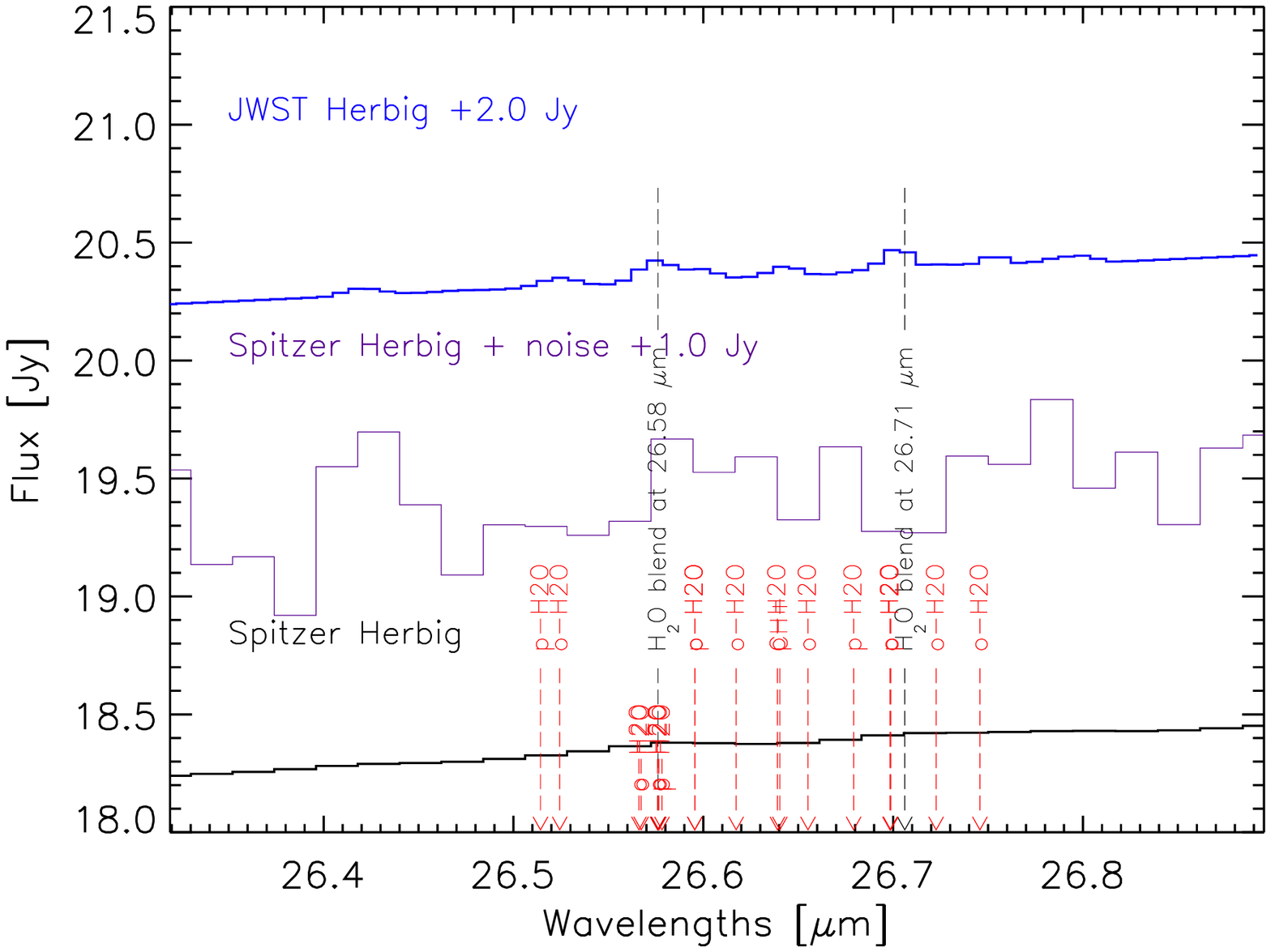}
\caption{Theoretical spectra convolved and re-binned at different resolutions (Spitzer IRS, $R$~=~600; JWST MIRI, $R$~=~3000 at 5.5~$\mu$m; VLT VISIR, $R$~=~25000), and zoom in the 26~$\mu$m region. The ranges adopted are the 
same as the instruments spectrographs. VISIR is ground based, and the main limits is due to atmospheric transmission, we plotted only the region in which the transmissivity is $\ge$~80\%. The Spitzer spectrum with noise is 
also plotted in magenta. The strong feature at 12.93~$\mu$m in the VISIR modeled spectrum of a T~Tauri star disk is due to H$_2$ ($J~=~7-6$, $v~=~0$). Water lines contributing to the blends at 26~$\mu$m are marked 
as vertical red arrows.}
\label{Comparison}
\end{figure*}

Theoretical spectra of the global mid-IR spectrum and some zoom on two regions (12.4~$\mu$m and 26.6~$\mu$m) convolved at the resolutions of Spitzer IRS \citep{houck}, 
JWST MIRI \citep[][]{clampin} and VISIR on VLT \citep[][]{lagage1} (see Appendix \ref{app}), are plotted in Fig.~\ref{Comparison}. 
The plot shows that the improvement due to higher spectral resolution will allow to detect unblended water transitions with larger line/continuum ratio.
The resolution of JWST/MIRI will improve the detection rate, but the gain is potentially still too low for Herbig stars, and higher resolution ($R~\gg~3000$) is needed.
The comparison between Herbig and T~Tauri disks shows that the features are stronger for the earlier type objects (see Fig.~\ref{Spitzer-Teff}), but the lines are also buried in a stronger continuum. The related stronger 
noise then makes the water lines even more difficult to be detected for Herbig PMS stars. T~Tauri disks show weaker water line fluxes, but the lower continuum translates into a lower noise and so into a higher detection rate. 
This is in agreement with the more successful detection of mid-IR water blends towards T~Tauri stars (about 50\%).

\subsection{Near-IR ro-vibrational water lines}

\begin{figure}[htpl!]
\centering
\includegraphics[width=0.48\textwidth]{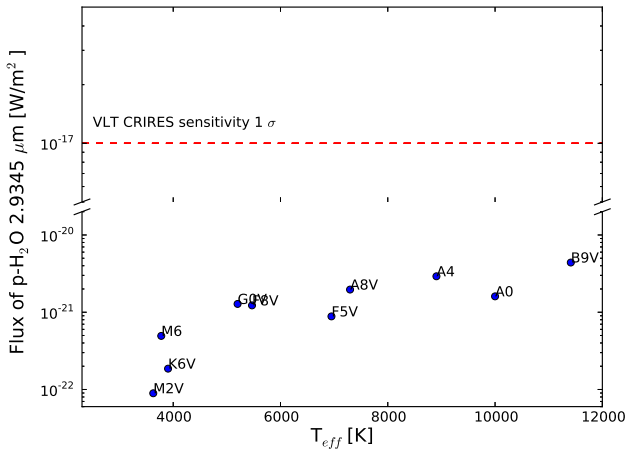}
\caption{p-H$_2$O 2.394~$\mu$m ro-vibrational line prediction from models with different central star. Typical sensitivity limits from the observations in \citet{fedele1} (10-40 minutes exposure) are marked as red dashed line.}
\label{CRIRES}
\end{figure}

As sanity check of the quality of our models, we computed the predicted line flux for a ro-vibrational water transition at 2.934~$\mu$m (Fig.~\ref{CRIRES}) targeted by \citet{fedele1}. Our models, in agreement with the
observations, predict that this ro-vibrational line should be undetectable in both T~Tauri and Herbig star disks, due to sensitivity limits of the CRIRES spectrograph on the VLT \citep{kaeufl}. 

\section{Discussion}\vspace{5mm}
\label{4}

In this work, we investigate a couple of explanations for the non detection of mid-IR water towards more massive PMSs, based on our modeling physics and capabilities. The real disks can be more complex than the parametrized 
models studied here. In the following subsections, we will address individually the possible explanations for the low water detection rate in Spitzer spectra of Herbig stars.

\subsection{Modeling assumptions}

We consider here parametrized disk structures with monotonic shape and radial mass distribution. Several previous studies pointed out that the stronger irradiation from a hotter central star should produce a puffed-up inner disk
\citep[e.g.][]{dullemond4}.  Observations towards some Herbig stars disks suggests that this scenario could be common \citep[e.g. HD~163296, HD~141569A, and HD~150193A; ][]{garufi}.
This change in the inner disk is able to cast a strong shadows, but apparently from hydrostatic modeling it is not enough to produce lower temperatures in the inner few au \citep{meijerink1}, and so mid-IR line fluxes should not
be affected.

Our code is able to compute hydrostatic equilibrium disks, but we are limited by the heating/cooling processes implemented and the results tend towards a maximum flaring angle \citep{meijerink1}, which is generally not 
seen in detailed fitting results of individual objects \citep[e.g. ][]{tilling,garufi1}. The hydrostatic assumption itself is a limitation since it suppresses winds in disks, and this aspect is in conflict with the theory on 
photoevaporation \citep[e.g. ][]{hollenbach3}.

The rockline displacement in disks around bright PMSs, is not large enough to explain the water non detections. Our models suggest that in order to suppress the mid-IR water spectra, it is necessary to open a gap larger than 
20~au or to lower the disk scale height below 10$^{-3}$~au in the inner disk within 0.5~AU. From a recent near-IR interferometry survey of Herbig stars, about 5\% of the objects are consistent with a gap larger than 20~au 
\citep{menu}.

\subsection{Intrinsic disk physics}

Following the outcome of the first series of models, we performed a parameter space exploration to understand if physical differences between low and high mass PMSs disks could be a candidate explanation for the non
detections. We found that indeed several properties are able to suppress the mid-IR water lines below the typical sensitivity range we considered (8$\times$10$^{-17}$-8$\times$10$^{-18}$~W/m$^2$): very low disk 
scale height, large gaps (see previous Section), and high dust opacity (due to high dust content or grain size). These results are in agreement with our previous results on T~Tauri stars. 

Then also the standard model we took as fiducial case of a typical disk surrounding an Herbig star, is very close to the sensitivity range. The observations of Herbig stars overplotted on the parameter 
space explored, shows that disks populate largely the region we explore, and several objects could have been in principle detected only if they were close enough. In \citet{pontoppidan1} we found that many of these 
targets were at distances larger than 220 pc. 
Again our model seems consistent with the observations.
From our results, we are not able to exclude intrinsic differences in disks around Herbig stars with respect lower mass PMS stars. However, observations from  \citet{acke2} conclude that 50\% of the Herbig stars analyzed 
show scale heights enlarged by a factor 2 or 3 with respect hydrostatic prediction. This strongly puffed up inner region is able to cast a shadow that makes the disk cooler, as for HD~95881 \citep{verhoeff}, possibly also for 
HD~144432 \citep{chen}, and finally HD~163296 and HD~150193A \citep{garufi}.
Then recent continuum radiative transfer modeling suggests that the mid-IR emitting regions of disks around different PMS stars (spanning from Brown dwarfs through Herbigs) are self-similar, in surface density, vertical 
distribution of dust and gas, and grain size distribution \citep{mulders1}.

\subsection{Instrumental noise}

\begin{figure}[htpl!]
\centering
\includegraphics[width=0.48\textwidth]{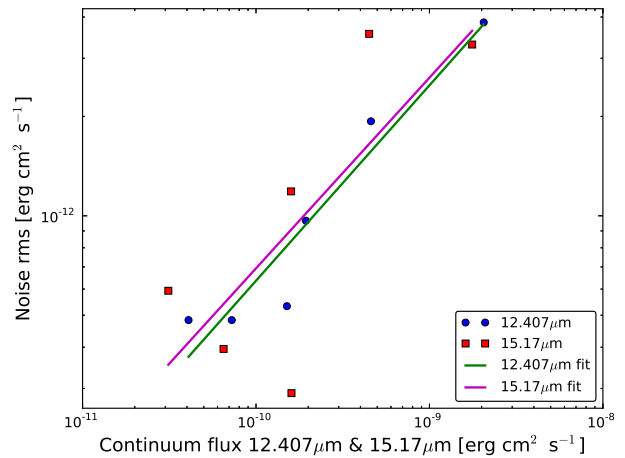}
\caption{Noise level versus continuum flux at the wavelengths of 12.407~$\mu$m (blue dots) and 15.17~$\mu$m (red squares). Overplotted are the fit to the data, green
for 12.407~$\mu$m and magenta for the 15.17~$\mu$m.}
\label{noise}
\end{figure}

Finally we take into account the instrumental noise, that is able to affect relevantly the detection rate towards bright PMSs disks.
Our study suggests that the low resolution and sensitivity of Spitzer IRS is the main cause of mid-IR water non detections in disks around Herbig stars. Mid-IR lines are buried in a stronger continuum, and the corresponding 
larger noise.
We consider here a linear relation between noise and continuum, that is a conservative criterium. Pattern-noise and fringing\footnote{http://irsa.ipac.caltech.edu/data/SPITZER/docs/irs/features/} produce systematically 
stronger noise at high signal levels (i.e. for stronger continuum, Fig.~\ref{noise}). This noise can be reduced adopting deeper integration or multiple integration cycles. The observations analysed in \citet{pontoppidan1} were 
deep enough for a relevant number of T~Tauri stars, but not for Herbig stars.

Higher spectral resolution is needed to investigate water and likely other weaker mid-IR lines in disks around early type PMS stars, overcoming noise and integration time problems. 
JWST/MIRI would unblend already several Spitzer water lines, and will allow a better line/continuum ratio, but this could still be not high enough for Herbig stars. Ground based facilities, with spectral resolution $>$~20000 
(like VISIR on VLT), are needed to improve the detection of water towards a wider sample of Herbig stars. However, ground-based observations will always be limited in wavelength coverage (and excitation temperatures) due to
atmospheric transmission and telluric lines, and would limit the detections to some energetic rotational transitions as the 12.4~$\mu$m blend observed in \citet{pontoppidan1}.

\section{Conclusions}
\label{5}

From the outcome of our model we provide a scenario that can explain the observations in the mid-IR towards Herbig star: we considered some limitations in our description of disks around Herbig stars, and we think that these 
would not be problematic for the correct physical description of the mid-IR spectroscopy. Observations have been performed up to now with facilities affected by low spectral resolution and lower sensitivity. These two elements 
limit any possible firm conclusion about the nature of disks around Herbig stars, that can be intrinsically different from the ones surrounding T~Tauri stars. Low $S/N$ is the most likely explanation for the low 
detection rates of mid-IR water lines around early type PMSs. This problem affects also partially later type PMSs: T~Tauri stars spectra are less noisy, however, mid-IR water line fluxes are lower, and the 
detection rate is only a factor two larger than for Herbig stars.
Better performing instruments, with resolution $>$ 3000 and longer integration times, are definitively needed to detect water successfully in the range 1-20~$\mu$m in protoplanetary disks.

\begin{acknowledgements}
The research leading to these results has received funding from the European Union Seventh Framework Programme FP7-2011 under grant agreement no 284405
\end{acknowledgements}

\bibliographystyle{../aa}
\bibliography{../Draft}

\appendix

\section{Sensitivity limits}
\label{app1}

Our analysis sample includes a selection of T~Tauri and Herbig Spitzer-IRS high resolution observations. BCD data products were retrieved from the Spitzer archive\footnote{http://irsa.ipac.caltech.edu/data/SPITZER/docs/spitzerdataarchives/}. 
The data were reduced with the c2d IRS science pipeline using the PSF extraction method \citep[see][for details]{lahuis2007b,lahuis2007PhD}, which provides higher S/N spectra then the standard tapered aperture extraction used to
populate the Spitzer archive. After extraction, additional defringing \citep{lahuis2003} was applied to remove any low-level fringe residuals.

Sensitivity limits were determined by calculating the line flux for synthetic lines with an amplitude equivalent to the 1-$\sigma$ noise level in the IRS spectrum at the respective blend wavelengths and a width comparable to the 
blend width at a spectral resolution of 600.

\section{Spectral convolution}
\label{app}

We applied Gaussian convolution and rebinning to the theoretical spectra from our models. For this task we used the ISO spectral analysis package \citep{kessler}, considering the three resolutions of Spitzer IRS LH/SH modules 
\citep[$R$~=~600;][]{houck}, VLT VISIR \citep[$R$~=~25000;][]{lagage1}. For JWST MIRI we split the range in four spectral windows: $R$~=~2800 for 7.63-11.71~$\mu$m, $R$~=~2500 for 11.71-13.59~$\mu$m, $R$~=~2300 for 13.59-18.14~$\mu$m,
$R$~=~1600 for 18.14-28.43~$\mu$m, accounting for the variable spectral resolution \citep[][]{clampin}. We add noise through a random generator according to 

\begin{equation}\label{(1)}
\centering
N = F_\mathrm{cont}\cdot \frac{U([0,1])}{100}
\end{equation}

$N$ is the computed noise, it is a function of the continuum flux ($F_\mathrm{cont}$) multiplied for a random generator between 0 and 1 ($U([0,1])$), and of the selected $S/N$ ratio, here we consider 100 as claimed in 
\citet{pontoppidan1}.

\section{Far UV spectra}
\label{app2}

FUV data are produced combining observations from IUE, FUSE, HST/STIS, HST/COS and when available HST/ACS archival data (Fig.~\ref{dt}). These data have been taken in multiple observations of the same objects, and have 
different resolution, integration time, spectral range and noise. Due to the complex dataset, the best spectra we use is an average performed through a sophisticated procedure described in detail in Dionatos et al in prep.

\begin{table*}
\caption{Far UV spectra data source, and number of epochs for each source.}
\centering
\begin{tabular}{|c|c|c|c|c|c|}
\hline\hline
Star & IUE & FUSE & STIS & COS & ACS \\ \hline
HD 100546 & 9 & 6 & 8 & 0 & 0 \\
HD 97048 & 12 & 0 & 0 & 0 & 0 \\
HD 163296 & 31 & 6 & 8 & 0 & 0 \\
HD 142666 & 7 & 2 & 0 & 0 & 0 \\
HD 135344B & 0 & 2 & 0 & 8 & 0 \\
RY Tau & 107 & 0 & 6 & 0 & 1 \\
RY Lup & 4 & 0 & 4 & 0 & 1 \\
DN Tau & 19 & 0 & 0 & 0 & 1 \\
CY Tau & 0 & 0 & 5 & 0 & 2 \\
DO Tau & 0 & 0 & 0 & 0 & 1 \\ \hline
\end{tabular}
\label{dt}
\end{table*}

\begin{figure*}[htpl!]
\centering
\includegraphics[width=0.4\textwidth]{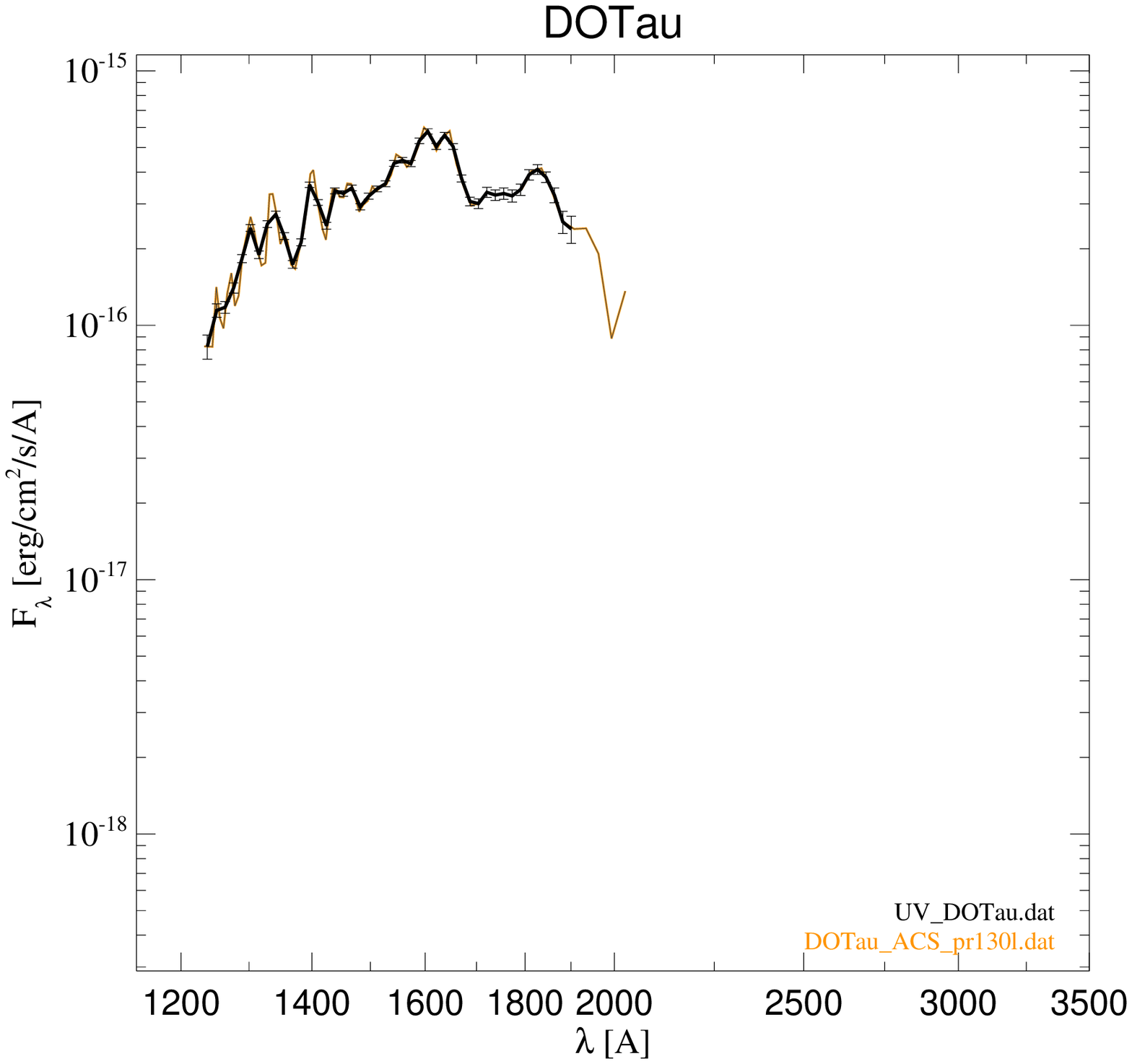}
\includegraphics[width=0.4\textwidth]{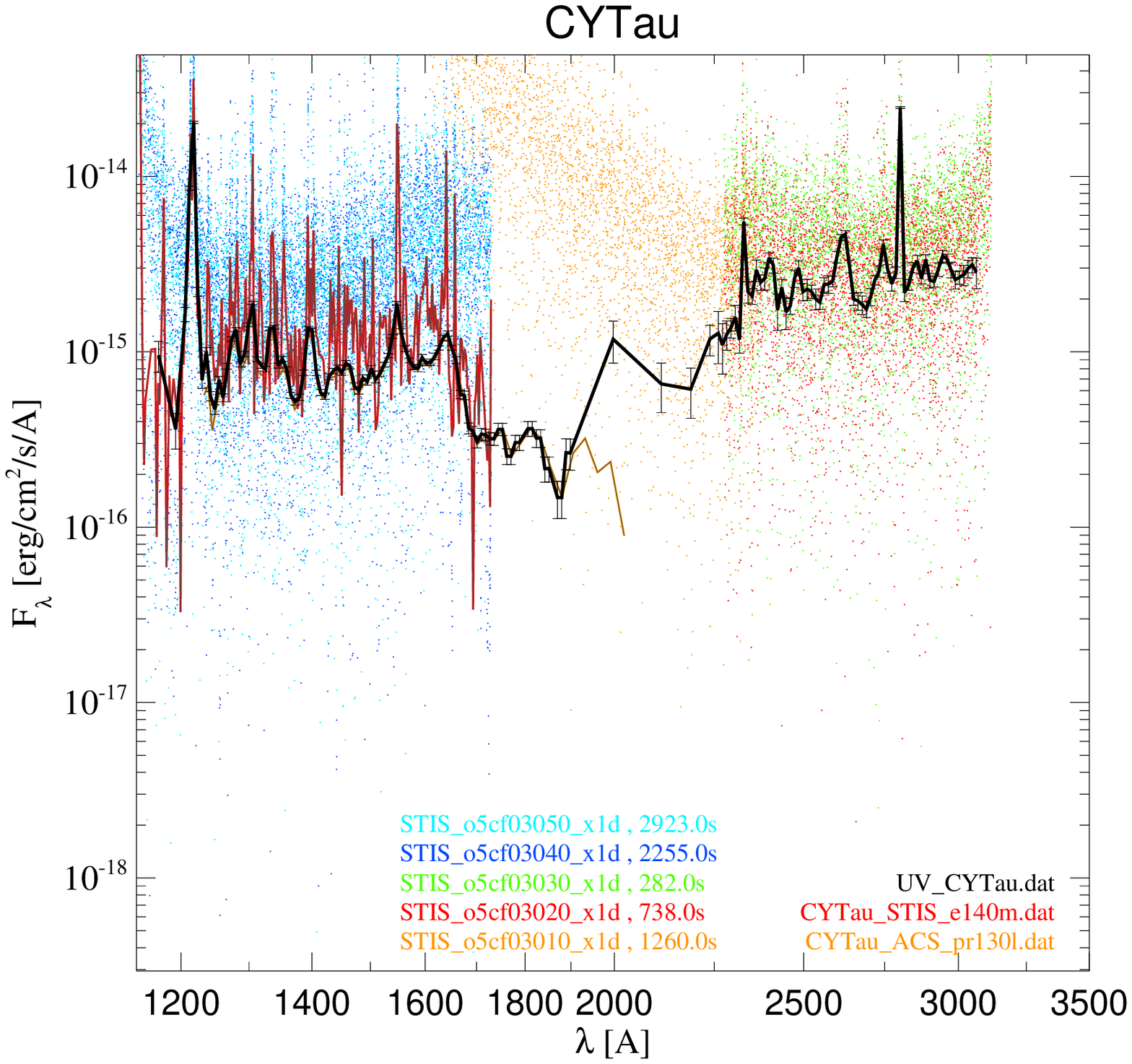}
\includegraphics[width=0.4\textwidth]{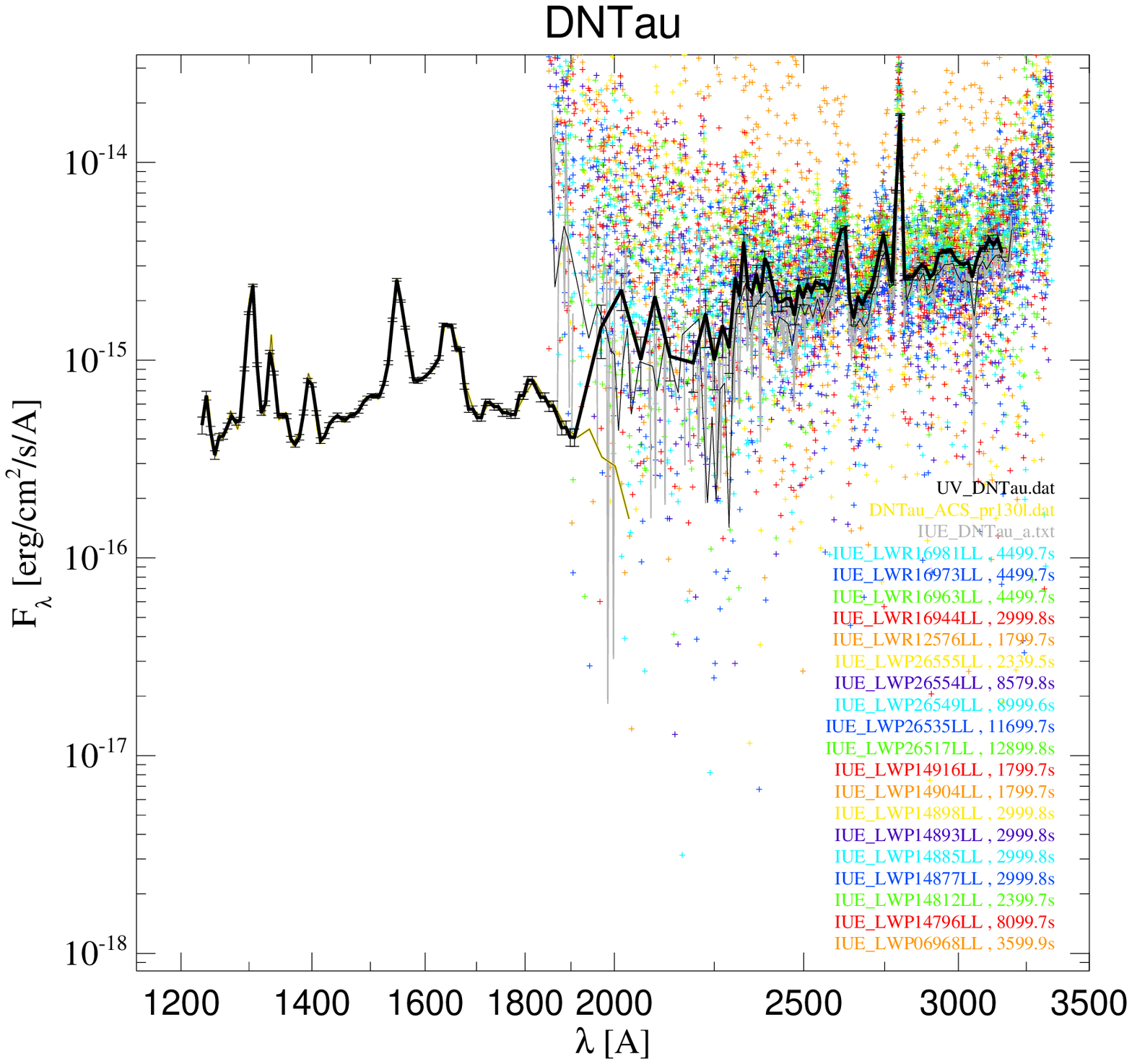}
\includegraphics[width=0.4\textwidth]{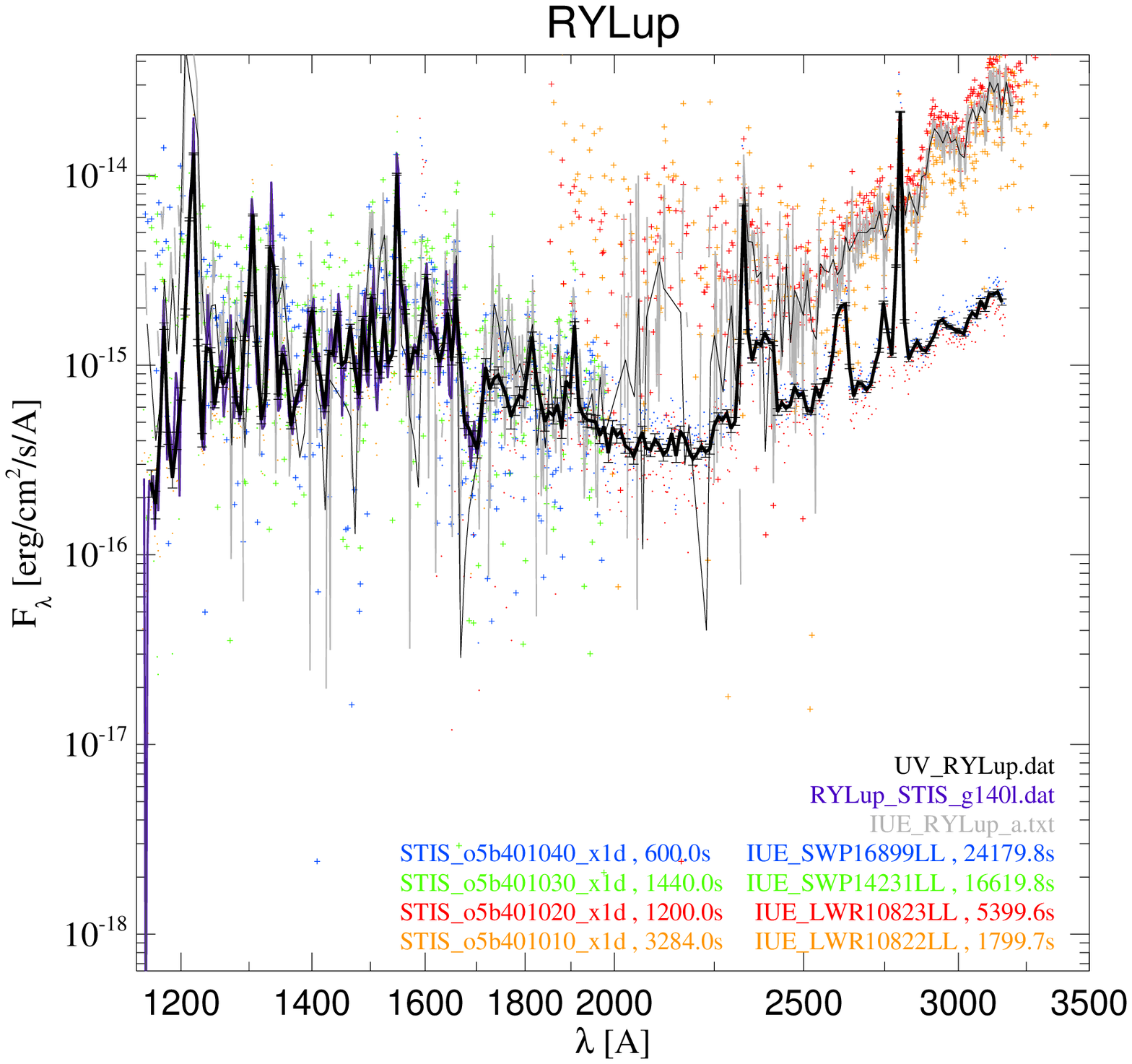}
\includegraphics[width=0.4\textwidth]{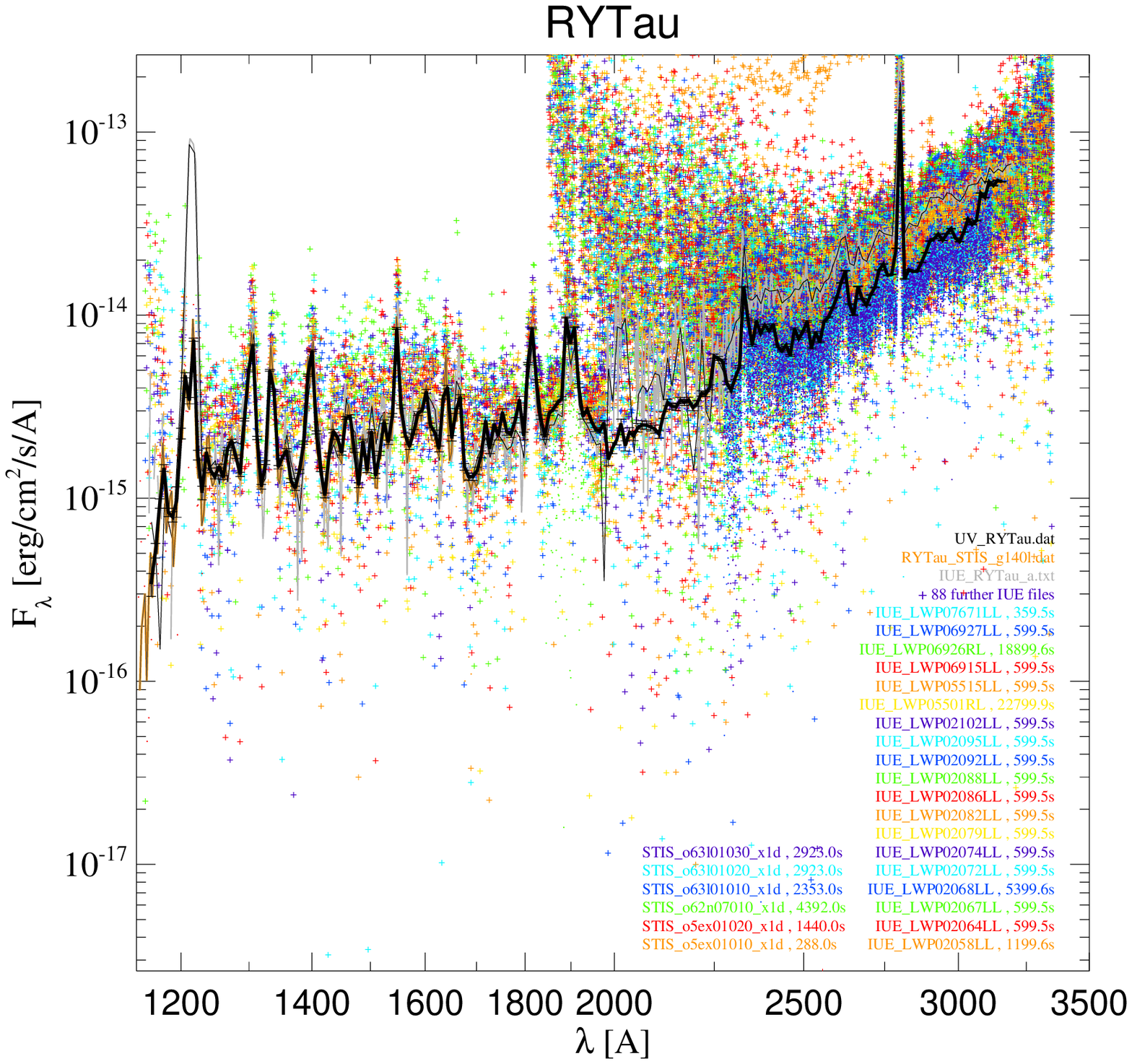}
\includegraphics[width=0.4\textwidth]{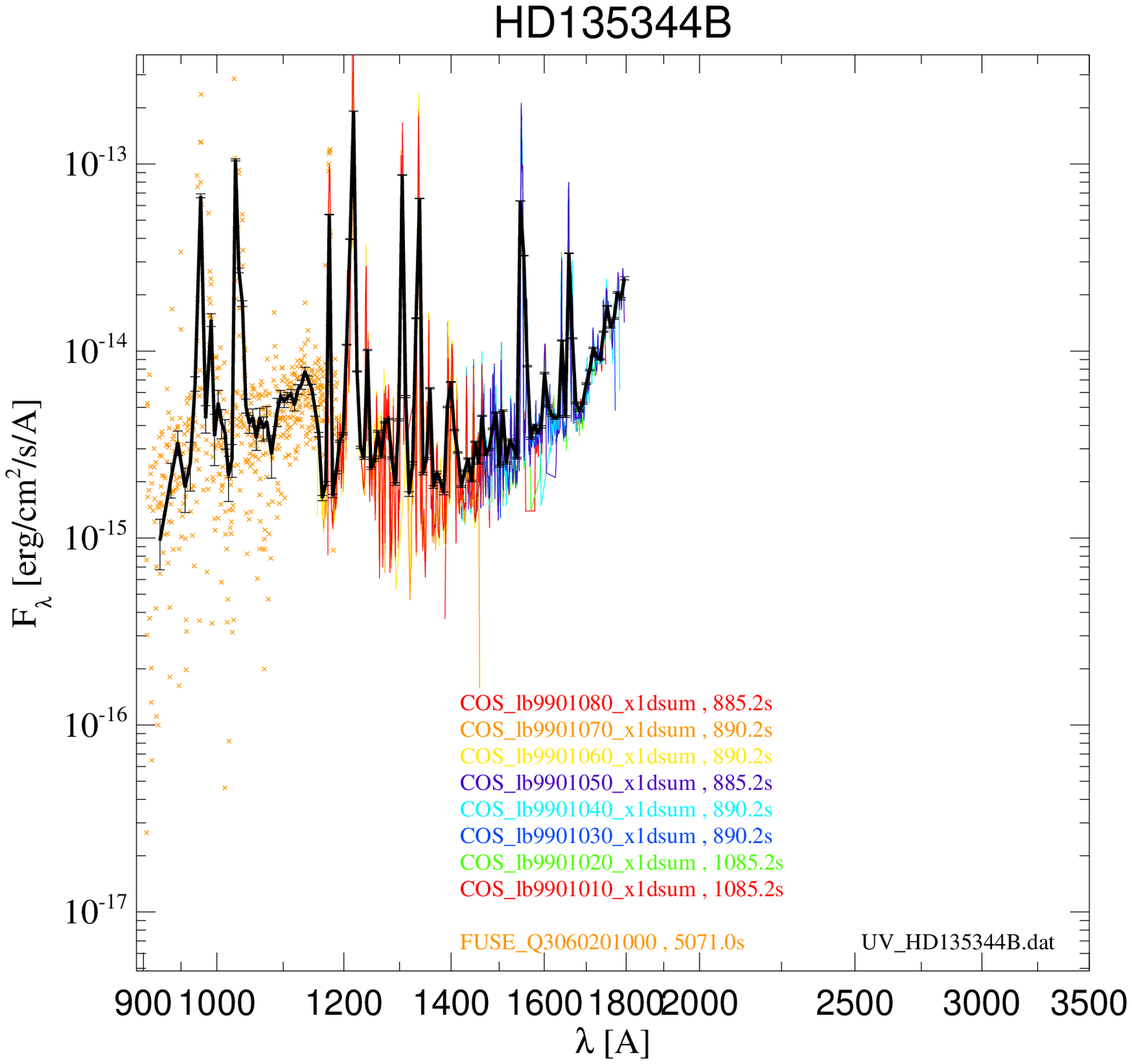}
\label{UVC}
\caption{Far UV component for each of the stars we modeled. The black line is the best fit of the multiple epochs of observations. Spectra data points are color coded, and the original files are labeled.}
\end{figure*}

\begin{figure*}[htpl!]
\centering
\includegraphics[width=0.4\textwidth]{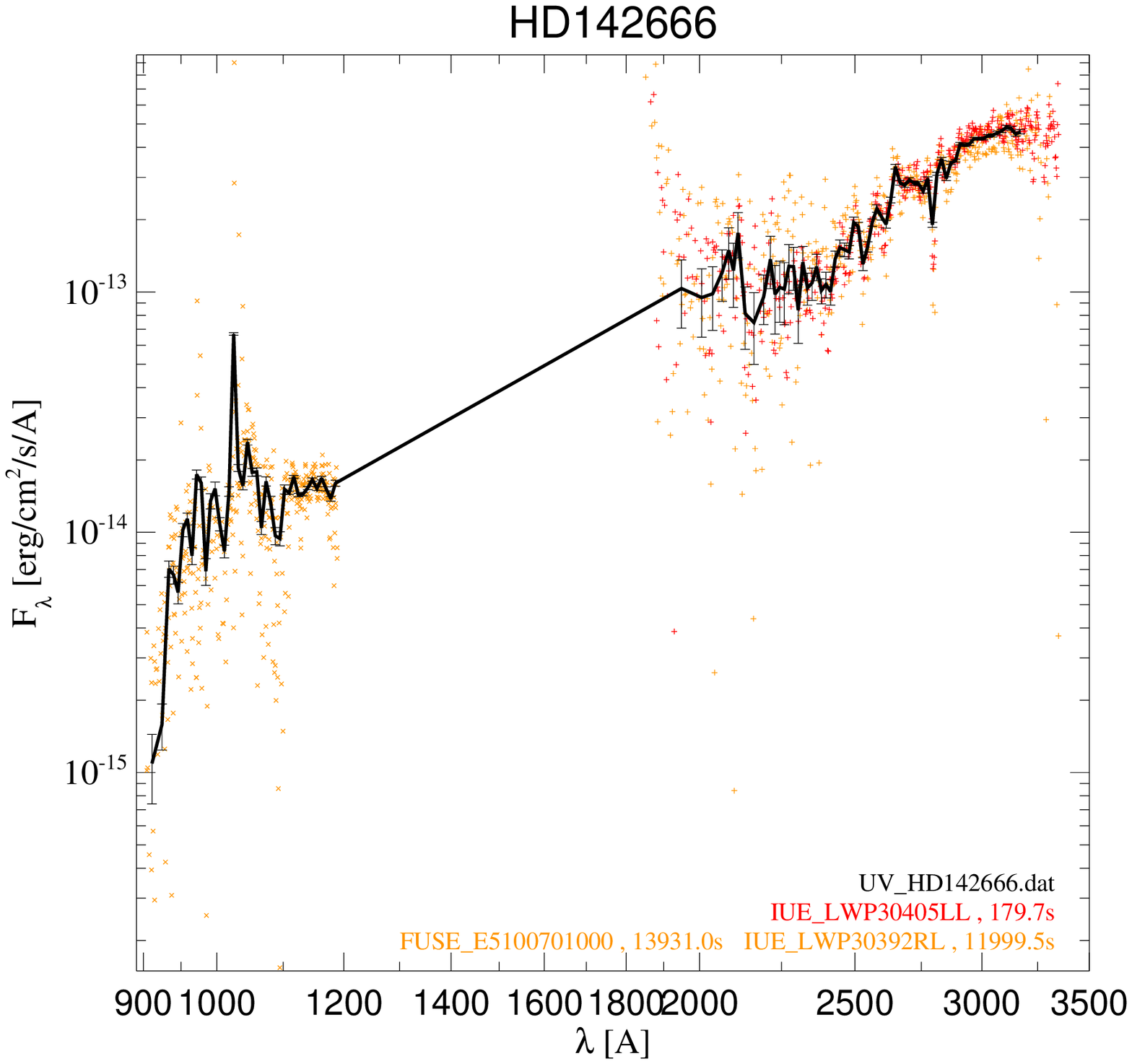}
\includegraphics[width=0.4\textwidth]{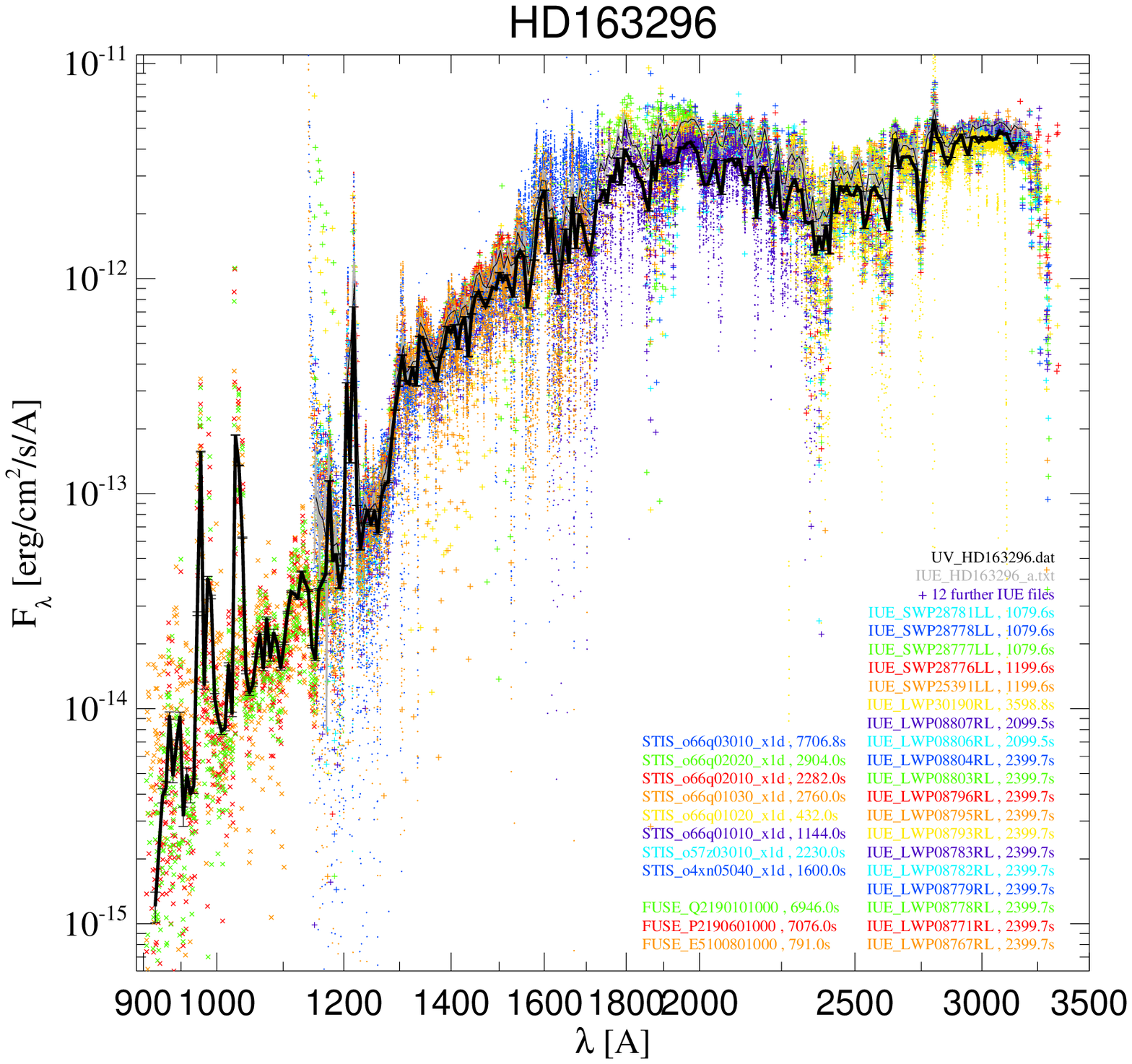}
\includegraphics[width=0.4\textwidth]{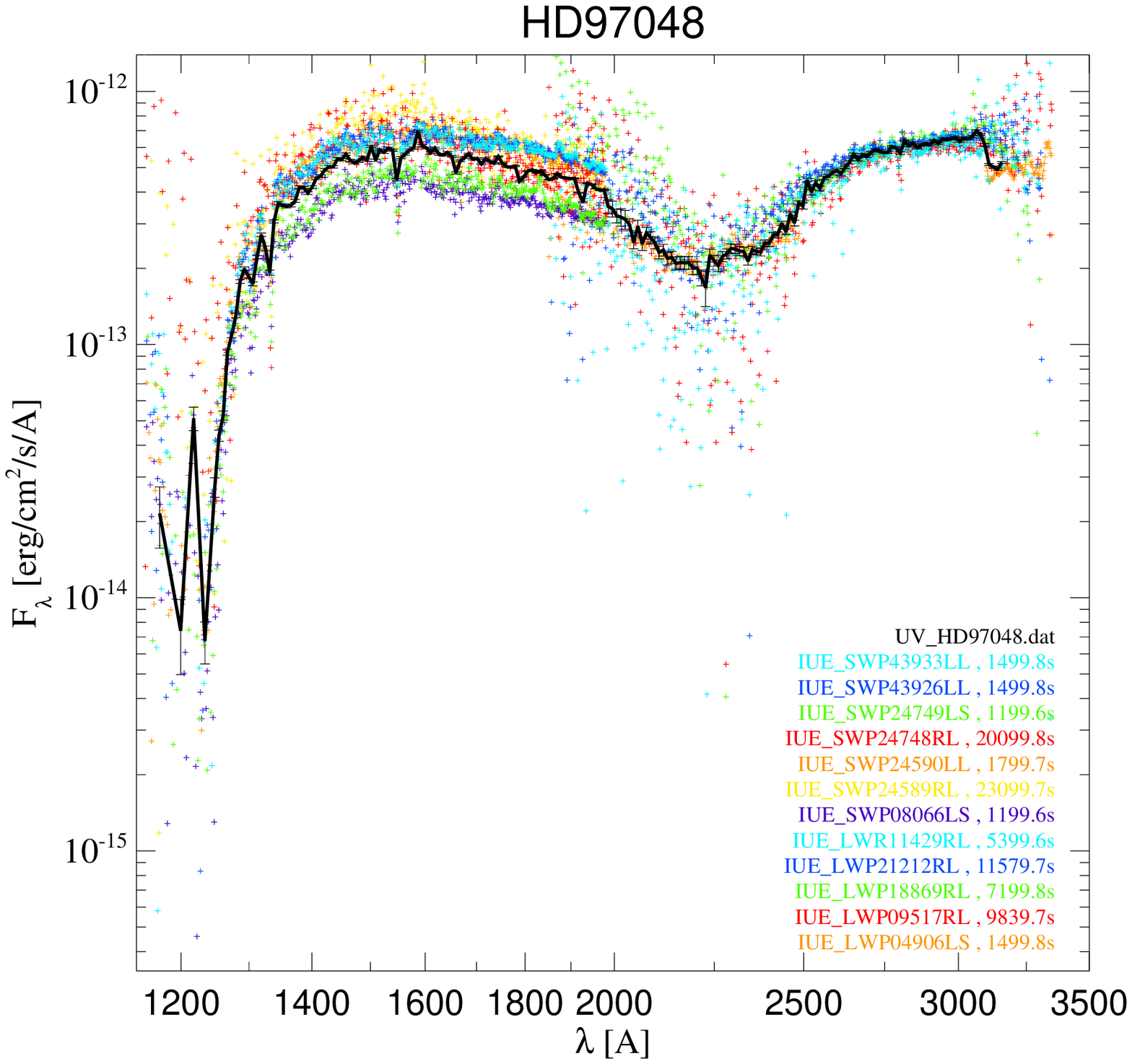}
\includegraphics[width=0.4\textwidth]{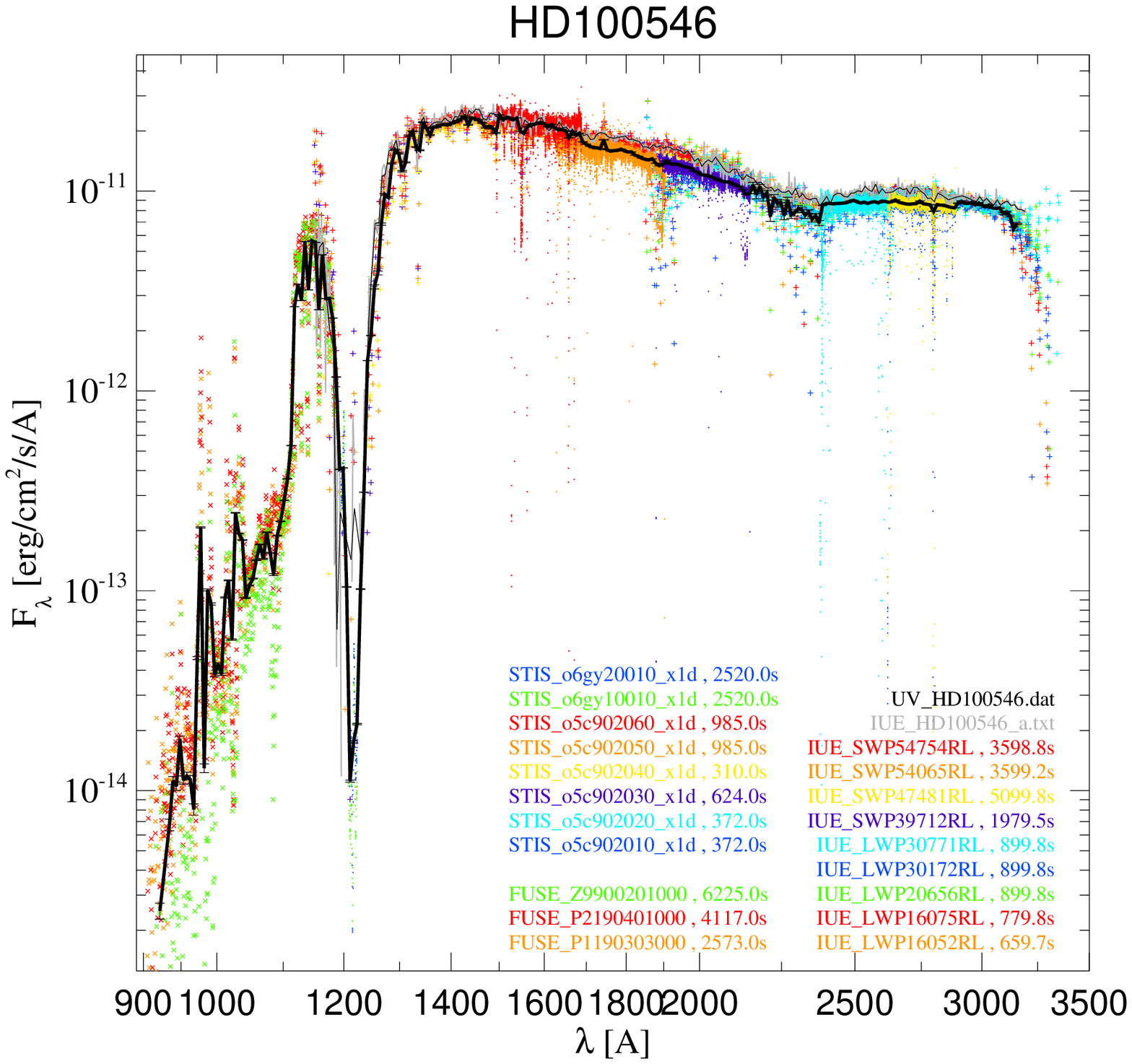}
\caption*{Fig.~\ref{UVC} Continued.}
\end{figure*}

\end{document}